\documentclass[pre,twocolumn]{revtex4}
\usepackage{graphicx}
\usepackage{amsmath,amssymb,amsthm,amsbsy}
\usepackage{latexsym}
\usepackage{amsfonts}
\usepackage{amssymb}

\font\tenbg=cmmib10 at 10pt

\def \rvecphi{{\hbox{\tenbg\char'036}}}

\font\tenbg=cmmib10 at 10pt

\def \rvecphi{{\hbox{\tenbg\char'036}}}

\begin{document}

\title{ Bunching
Instability of Rotating Relativistic
Electron Layers and Coherent Synchrotron
Radiation}

\author{Bjoern S. Schmekel}
\affiliation{Department of Physics,
Cornell University, Ithaca, New York 14853}
\email{bss28@cornell.edu}

\author{Richard V.E. Lovelace, and Ira M. Wasserman}
\affiliation{Department of Astronomy,
Cornell University, Ithaca, New York 14853}
\email{rvl1@cornell.edu, ira@astro.cornell.edu}

%\author{Ira M. Wasserman}
%\affiliation{Department of Astronomy,
%Cornell University, Ithaca, New York 14853}
%\email{ira@astro.cornell.edu}

\begin{abstract}
We study the stability of a collisionless,
relativistic, finite-strength,
cylindrical layer of charged particles
in free space by solving the
linearized Vlasov-Maxwell equations and compute
the power of the emitted electromagnetic waves.
   The layer is rotating in an external
magnetic field parallel to the layer.
This system is of interest to
understanding the high brightness
temperature of pulsars which cannot
be explained by an incoherent
radiation mechanism.
    Coherent synchrotron radiation
has also been observed recently in
bunch compressors used in particle accelerators.
   We consider equilibrium layers with a
`thermal' energy spread and therefore
a non-zero radial thickness.
   The particles interact with their retarded electromagnetic
self-fields.
   The effect of the betatron oscillations is retained.
A short azimuthal wavelength
instability is found which causes a
modulation of the charge and current densities.
   The growth rate is found to be
an increasing function of the
azimuthal wavenumber, a decreasing function
of the Lorentz factor, and proportional
to the square root of the total
number of electrons.
     We argue that the growth
of the unstable perturbation saturates
when the trapping frequency of
electrons in the wave becomes
comparable to the growth rate.
        Owing to this saturation we can
predict the radiation spectrum
for a given set of parameters.
Our predicted brightness temperatures are proportional
to the square of the number of particles and scale by
the inverse five-third power of the
azimuthal wavenumber which is in rough accord
with the observed spectra of radio pulsars.

\end{abstract}

\maketitle

\section{Introduction}

The high brightness temperatures
of the radio emission of pulsars
($T_B \gg 10^{12} $K) implies
a coherent emission mechanism
\citep{Gold1968,Gold1969,GoldreichKeeley1971,
ManchesterTaylor1977,Melrose1991} and some part of
the radio emission of extragalactic jets
may be coherent \citep{GBKRL95}.
Recently, coherent synchrotron
radiation (CSR) has  been
observed in bunch compressors
\citep{Loos2002,Byrd2002,Kuske2003} which are a
crucial part of future particle accelerators.
       When a relativistic beam
of electrons interacts with its own
synchrotron radiation the beam may become modulated.
If the wavelength of the modulation is less than the
wavelength of the emitted
radiation, a linear instability may occur
which leads to exponential
growth of the modulation amplitude.
The coherent synchrotron
instability of relativistic electron rings
and beams has been investigated theoretically by
\cite{GoldreichKeeley1971,
HeifetsStupakov2001,Stupakov2002,Heifets2001,Byrd2003}.
Goldreich and Keeley analyzed the stability of a ring
of monoenergetic relativistic electrons
which were assumed to move on a circle of fixed
radius. Electrons of the ring gain or lose
energy owing to the tangential electromagnetic force and at
the same time generate the electromagnetic
field. \cite{Uhm1985} analyzed the stability of a relativistic
electron ring enclosed by a conducting beam pipe
in an external betatron magnetic field. A distribution function
with a spread in the canonical momentum was chosen for
their analysis. For simplicity the effect of the betatron oscillations
was not included in their treatment. They find a resistive wall
instability and a
negative mass instability. Furthermore, they find an instability
which can perturb the surface of the beam. \cite{Heifets2001}
analyzed the stability of a ring of relativistic electrons
in free space including a small energy spread which gives a range
of radii such that particles on the inner orbits can pass
particles on outer orbits. \cite{Byrd2003} has developed a similar
model which includes the effects of the conducting beam pipe.
Numerical simulations by \cite{Venturini2003} show the burst-like nature
of the coherent synchrotron radiation.

The present work analyzes the linear
stability of a cylindrical, collisionless, relativistic
electron (or positron) layer or E-layer \citep{Christofilos1958}.
Particle densities in pulsar magnetospheres are very
low, of order the Goldreich-Julian charge density
$n_{GJ}={\bf\Omega\cdot B}/2\pi ce\sim 
10^{11}s\,{\rm cm^{-3}}(B/10^{12}\,{\rm G})(R/r)^3
[P({\rm sec})]^{-1}$ at radius $r>R$, where $R$ is the
stellar radius, $B=10^{12}B_{12}\,{\rm G}$ is the
surface field strength, and $P$ is the rotational period;
thus, the magnetospheric plasma is collisionless to an
excellent approximation \cite{Julian1969}.
          The particles in the layer
have a finite `temperature' and
thus a range of radii so that
the limitation of the Goldreich and
Keeley model is overcome.
Although we allow a spread in energies, we assume that it
is small, so the charge layer is also thin; efficient
radiation losses are probably sufficient to maintain rather
low energy spreads in a pulsar magnetosphere, although
the precise size of the spread is still not entirely
certain.
          Viewed from a moving frame the E-layer is
a rotating beam. The system is sufficiently simple
that it is relevant to electron
flows in pulsar magnetospheres (cf. \cite{Arons2004}).
          The analysis involves solving
the relativistic Vlasov equation using
the full set of Maxwell's equations and computing
the saturation amplitude due to trapping. The latter
allows us to calculate the energy loss due to
coherent radiation.

          In \S 2 we describe the considered Vlasov equilibria.
The first type of equilibrium (a) is formed by
electrons (or positrons) moving perpendicular to
a uniform magnetic field in the
$z-$direction so as to form a thin
cylindrical layer referred
to as an E-layer.
         The second type of equilibrium
(b) is formed by electrons moving almost parallel
to an external toroidal magnetic field and
also forming a cylindrical layer.
          \S 3 describes the method of solving the
linearized Vlasov equation which involves
integrating the perturbation force along the
unperturbed orbits of  the equilibrium.
           In \S 4, we derive the dispersion relation for
linear perturbations  for the case
of a radially thin E-layer and
zero wavenumber in the axial direction, $k_z=0$.
          We find that there is in general a short wavelength
instability.
            In \S 5 we analyze the nonlinear saturation
of the wave growth due to trapping of the
electrons in the potential wells of the
wave.
          This saturation allows the
calculation of the actual spectrum of
coherent synchrotron radiation.
         In \S 6, we derive the dispersion relation
for linear perturbations of a thin E-layer
including a finite axial wavenumber.
          The linear growth is found to occur
only for small values of the axial wavenumber.
          The nonlinear saturation due to trapping
is similar to that for the case where $k_z=0$.
In \S 7 we consider the effect of the thickness
of the layer more thoroughly and include
the betatron oscillations.
           \S 8 discusses the apparent brightness
temperatures for the saturated coherent
synchrotron emission.
\S 9 discusses some implications on
particle accelerator physics.
\S 10 gives conclusions of this work.

\section{Equilibria}

\subsection{Configuration a}

We first discuss the Vlasov equilibrium
for an axisymmetric, long, thin cylindrical
layer of relativistic electrons where
the electron motion is almost perpendicular
to the magnetic field.
This is shown in Figure 1a.  The case where
the electron motion is almost parallel to the
magnetic field is discussed below.
           The equilibrium has $\partial/\partial t=0,
~\partial/\partial \phi =0,~$ and $\partial/\partial z=0$.
           The configuration is
close to the non-neutral Astron E-layer of \cite{Christofilos1958}.
            The equilibrium distribution
function $f^0$ can be taken to be
an arbitrary non-negative function of the
constants of motion, the Hamiltonian,
\begin{eqnarray}
H \equiv \left (m_e^2 +p_r^2+ p_{\phi}^2
+ p_z^2 \right )^{1/2} - e\Phi^s(r)~,
\end{eqnarray}
and the canonical angular momentum,
\begin{eqnarray}
P_{\phi} \equiv r p_{\phi} - er A_{\phi}(r)~,
\end{eqnarray}
where $A_\phi =A_\phi^{e} + A_\phi^{s}$
is the total (external plus self) vector potential, $\Phi^s$
is the self electrostatic potential,
$m_e$ is the electron rest mass, $-e$ is
its charge, and the units are such that $c=1.$
          Here, the external magnetic field is assumed
to be uniform, ${\bf B}^e=B^e_z\hat{\bf z}$, with
$A_\phi^{e}=rB^e_z/2$, and $B^e_z>0$.
Thus we have $f^0 =f^0(H,P_\phi)$.
We consider the distribution function
\begin{eqnarray}
f^0 =K\delta(P_\phi-P_0)\exp\big[-H/T\big]~,
\label{equi}
\end{eqnarray}
where $K$, $P_0$, and $T$ are constants (see
for example \cite{Davidson1974}).
The temperature $T$ in energy units is assumed
sufficiently small that the fractional radial thickness of
the layer is small compared with unity.
Note that a Lorentz transformation in the $z-$direction
gives a rotating electron beam.

%%%%%%%%%%%%%%%%%%%%%%%%%%%%%%%%%%%%%%%%%%
\begin{figure}
\includegraphics[width=3.5in]{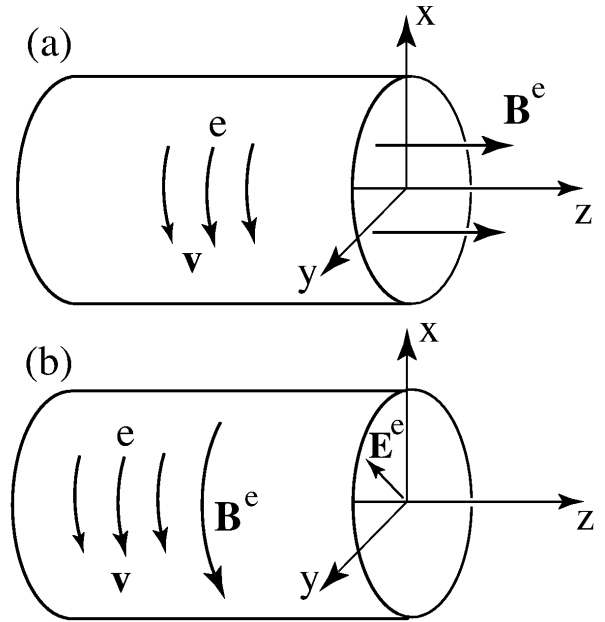}
\caption{
Geometry of relativistic E-layer in
({\bf a}) for the case of a uniform
external axial magnetic field, and in ({\bf b})
for an external toroidal magnetic field
with an external radial electric field.
}
\end{figure}
%%%%%%%%%%%%%%%%%%%%%%%%%%%%%%%%%%%%%%%%%

The equations for the self-fields are
\begin{eqnarray}
\frac{1}{r} \frac{d}{dr}
\left(r {d\Phi^s \over d r}
\right) = 4\pi e \int d^3 p~ f^0(H,P_{\phi})~,
\label{selfphi}
\end{eqnarray}
\begin{eqnarray}
\frac{d}{d r}\left( \frac{1}{r}
\frac{d( rA_\phi^{s})}{d r}\right)  =
4\pi e
\int d^3 p~ v_{\phi}~ f^0(H,P_{\phi})~,
\label{selfA}
\end{eqnarray}
where $v_\phi = (P_\phi/r +eA_\phi)/H$.

Owing to the small radial thickness of the layer,
we can expand radially near $r_0$
\begin{eqnarray}
\left[{P_\phi\over r}
+eA_\phi(r)\right]^2\!\!\!=\left[{P_\phi\over
r_0}\!\!+eA_\phi(r_0)\right]^2 \!\!+\delta r D_1 \!\!+\!{1 \over 2}
\delta r^2 D_2 ,
\label{expandp}
\end{eqnarray}
where $D_1$, $D_2$ are the derivatives evaluated at $r_0$,
and $\delta r \equiv r-r_0$ with $(\delta r/r_0)^2 \ll 1$.
We choose $r_0$ so as to eliminate the term linear in $\delta r$.
Thus,
\begin{eqnarray}
H=H_0 -e\Phi^s(r_0) +{1 \over 2 H_0}\big(p_r^2+p_z^2
+ H_0^2 ~\omega_{\beta r}^2~ \delta r^2\big)~,
\label{expandH}
\end{eqnarray}
where $\omega_{\beta r}$ is the radial betatron frequency,
and
\begin{eqnarray}
H_0
& \equiv&
m_e\left\{1+\left[{P_\phi\over r_0}+
eA_\phi(r_0)\right]^2\right\}^{1/2}~, \\ \nonumber
\gamma_0
& \equiv &
{H_0\over m_e}~, \\ \nonumber
v_{\phi0}
&\equiv&
{1\over H_0}\left[{P_\phi \over
r_0}+eA_\phi(r_0)\right]~.
\end{eqnarray}
          We assume $\gamma_0^2 \gg 1$ and $v_{\phi0} >0$ so
that $v_{\phi0} =  1-1/(2\gamma_0^2)$ to a good
approximation.
The ``median radius'' $r_0$ is determined by the condition
$$
{D_1\over 2 H_0}  -e ~{d\Phi^s\over dr}\bigg|_{r_0}=0~,
$$
or
\begin{equation}
{1\over H_0}\left({P_\phi\over r_0}+eA_\phi\right)
\left(-{P_\phi\over r_0^2}
+e~{dA_\phi\over dr}\right)\bigg|_{r_0}
=e~ {d\Phi^s\over dr}\bigg|_{r_0} .
\label{constraint}
\end{equation}
To a good approximation,
$$
r_0 ={m_e\gamma_0 v_{\phi0} \over (1-2\zeta)eB^e_z}\approx
{m_e\gamma_0  \over eB^e_z} (1+2\zeta ) \quad {\rm or}
$$
\begin{equation}
r_0^2=
{2P_\phi \over e B^e_z}[1+3\zeta+{\cal O}(\zeta^2)]~.
\label{r0}
\end{equation}
Here,
\begin{eqnarray}
\zeta\equiv-~{B_z^{s}(r_0)\over B^e_z}~,
\quad \quad{\rm with}\quad \quad
\zeta^2 \ll 1~,
\label{zeta1}
\end{eqnarray}
is the field-reversal parameter of Christofilos.
          For a radially thin E-layer of axial
length $L$ consisting of a total
number of electrons $N$, the surface density
of electrons is $\sigma =N/(2\pi r_0 L)$ and the surface
current density is $-ev_\phi \sigma$.
          Because $B_z^s(r_0)$ is one-half the full change of
the self-magnetic field across the layer, we have
$\zeta = r_e  N/(\gamma L)$, where $r_e=e^2/(mc^2)$ is
the classical electron radius.
           Notice that $N$, $\zeta$, and $\gamma L$ are
invariants under a Lorentz transform in the $z-$direction.

             The radial betatron frequency $\omega_{\beta r}$ is
given by
\begin{eqnarray}
H_0^2~\omega_{\beta r}^2={D_2\over 2} -{D_1^2 \over 4 H_0^2}
-H_0e{d^2\Phi^s \over dr^2}~.
\end{eqnarray}
          Using Eq. (\ref{constraint}) gives
$$
\omega_{\beta r}^2 ={1-4\zeta \over 1-2\zeta}~
{v_{\phi0}^2\over r_0^2}~
+{e  v_{\phi0}\over \gamma_0 m_e}~
{d^2 A_\phi \over dr^2}\bigg|_{r_0}\!
-{e\over \gamma_0 m_e}~{d^2\Phi^s \over dr^2}\bigg|_{r_0}
$$
\begin{equation}
\approx
{1-2\zeta\over r_0^2}  -
{\sqrt{2/\pi}~\zeta \over   r_0\Delta r \gamma^2}~.
\label{omega2betar}
\end{equation}
The term $\propto 1/\Delta r$ is the
sum of the defocusing  self-electric force
and the smaller focusing self-magnetic force.
          For the layer to be radially confined we need
to have $\zeta <\sqrt{\pi/2}~\gamma^2(\Delta r/r_0)$.
         For $\zeta \ll \gamma^2(\Delta r/r_0)$ and
$\zeta^2 \ll 1$, we have $\omega_{\beta r} = 1/r_0$
to a good approximation.

The number density follows from Eq. (\ref{equi}),
$$
n \approx n_0\exp\left(-{\delta r^2\over 2\Delta r^2}\right)
~~{\rm where}~~ \Delta r \equiv
\left({T \over H_0\omega_{\beta r}^2}\right)^{1/2}
$$
\begin{equation}
{\rm or} ~~ {\Delta r^2 \over r_0^2} \simeq
{v_{th}^2 \over  1-2\zeta -\sqrt{2/\pi}~
\zeta (r_0/\Delta r)/\gamma^{2} }
\end{equation}
where
\begin{eqnarray}
        v_{th}
\equiv \left({T \over \gamma_0 m_e}\right)^{1/2}  
\end{eqnarray}
and
\begin{eqnarray}
n_0=2\pi K H_0 T r_0^{-1} \exp \left ( - \frac{H_0 - e \Phi^s(r_0)}{T} \right ) ~.
\end{eqnarray}
           As mentioned we assume the layer to be radially
thin with $(\Delta r/r_0)^2 \ll 1$.
          Consequently, equations (\ref{selfphi}) and
(\ref{selfA}) become
\begin{eqnarray}
{ d^2 \Phi^s \over dr^2} & \approx &4\pi e n_0 ~
\exp\left(-{\delta r^2 \over 2\Delta r^2}\right)~,
\quad \quad  \nonumber \\
{ d^2 A_\phi^{s} \over dr^2} & \approx &4\pi en_0~ v_{\phi0}~
\exp\left(-{\delta r^2 \over 2\Delta r^2}\right)~.
\end{eqnarray}
          Thus we obtain
\begin{eqnarray}
          \zeta ={-B^{s}_z(r_0)\over B_z^e}={4\pi e n_0 v_{\phi0}\Delta r
\sqrt{\pi/2} \over B_z^e}
\end{eqnarray}
           The equilibrium is thus seen to be determined by
three  parameters,
\begin{eqnarray}
\zeta^2~,\quad v_{th}^2~,\quad {\rm and}\quad 1/\gamma_0^2~,
\end{eqnarray}
which are all small compared with unity.

\subsubsection{Equilibrium Orbits}

           From the Hamiltonian of Eq. (\ref{expandH}) we have
\begin{equation}
\!{d^2 \delta r \over dt^2}\!=
-   \omega_{\beta r}^2\delta r,~\rightarrow~
        \delta r(t^\prime)\!=
\delta r_i\sin[\omega_{\beta r}(t^\prime-t)+\varphi]~,
\label{rorbit}
\end{equation}
where $r-r_0=\delta  r_i \sin\varphi$.
For future use we express  the orbit so that
${\bf r}(t^\prime=t) ={\bf r}$, where $({\bf r},t)$
is the point of observation.
Also, we have
\begin{eqnarray}
{d\phi \over dt} ={ P_\phi +e r A_\phi(r) \over
m_e \gamma r^2} = \dot{\phi}(r_0) +
{d\dot{\phi} \over dr}\bigg|_{r_0} \delta r +..~,
\end{eqnarray}
so  that
\begin{eqnarray}
\phi(t^\prime)=\phi +(t^\prime -t)\dot{\phi}_0
+{1\over \omega_{\beta r}}
{\partial \dot{\phi}_0 \over \partial r}\bigg|_{r_0}\times
\nonumber \\
\bigg\{-\delta r_i\cos[\omega_{\beta r}(t^\prime-t)+
\varphi]+\delta r_i\cos(\varphi)\bigg\}~,
\label{phiorbit}
\end{eqnarray}
where $\partial \dot{\phi}/\partial r |_{r_0}
= -\dot{\phi}_0/r_0$.
            For $\zeta \ll \gamma^2(\Delta r/r_0)$ and
$\zeta^2 \ll 1$, we have
$\partial \dot{\phi}/\partial r|_{r_0}/\omega_{\beta r} = -1/r_0$
to a good approximation.
Because the E-layer is uniform in the $z-$direction,
\begin{eqnarray}
z(t^\prime)=z+(t^\prime -t)  v_z~.
\label{eqorbitz}
\end{eqnarray}
The orbits are necessary for the stability analysis.

%%%%%%%%%%%%%%%%%%%%%%%%%%%%%%%%%%%%%%%%%%%%%%%%%%%%%%%%%
\subsection{Configuration b}

          Here, we describe a Vlasov equilibrium
for an axisymmetric, long, thin cylindrical
layer of relativistic electrons where the
electron motion is almost parallel to the
magnetic field.
            The equilibrium distribution
function $f^0$ is again taken to be
given by Eq. (\ref{equi}) in terms of
the Hamiltonian,
$H$, and the canonical angular momentum,
$P_{\phi} \equiv r p_{\phi} - er A_{\phi}(r),$
where $A_\phi =  A_\phi^{s}$.
          We make the same  assumptions as
above, $\gamma^2 \gg 1$,
$T/(m_e\gamma) \ll 1$, and $\Delta r^2/r_0^2 \ll 1$.
          In this case
there is no external $B_z$ field.
          Instead, we include an external toroidal magnetic
field $B_\phi^e$ with corresponding vector
potential $A_z^e$ and an external electric
field ${\bf E}^e$ with potential $\Phi^e$.
         The fields ${\bf B}^e$ and ${\bf E}^e$
correspond to the magnetic and electric fields of
a distant,  charged, current-carrying flow
along the axis.  Thus,  $|{ E}^e_r| <|{B}_\phi^e|$.
           The considered external field is of course
just one of a variety of fields which give
electron motion almost parallel with the
magnetic field.
          Note also that the distribution function
is restricted in the respect that it does
not include a dependence on the canonical
momentum in the $z-$direction $P_z=m_e\gamma v_z
-eA_z$.

          The distribution function (3) gives
$J_z=0$ so that there is no toroidal self
magnetic field.
          Thus the self-potentials in this case are
also given by equations (\ref{selfphi}) and (\ref{selfA}).
           Equations (\ref{expandp}) - (\ref{constraint})
are also applicable
with the replacement of $\Phi^s$ by the
total potential $\Phi$.
          In place of Eq. (\ref{r0}) we find
\begin{equation}
r_0= {m_e\gamma v_\phi^2 \over
(1-2\zeta)e E_r^e(r_0)}\approx
{m_e\gamma\over e E_r^e(r_0)}(1+2\zeta)~,
\label{r0confII}
\end{equation}
        where $
\zeta \equiv  {B_z^s(r_0) / E_r^s(r_0)}.$
          We again have
$\zeta = r_e  N/(\gamma L)$, where $r_e=e^2/(mc^2)$ is
the classical electron radius and $L$ is the
axial length of the layer.
          Because $d^2 \Phi^e/dr^2 =-(1/r)d\Phi^e/dr$,
the radial betatron frequency is again
given by Eq. (\ref{omega2betar}) (with $\Phi$ now
the total potential) so that the orbits
given in \S 2.1.1 also apply in this case.
          The electron motion is almost parallel to
the magnetic field in that $(B_z^s/B_\phi^e)^2 =
\zeta^2 (E_r^e/B_\phi^e)^2 < \zeta^2 \ll 1$.
          Notice that Eq. (\ref{r0confII}) for $r_0$ is formal in
the respect that $E_r^e \propto 1/r$.
          Therefore, $r_0$ is in fact arbitrary
in this case.
          Because the wavelengths of the unstable
modes are found to be small compared with
$r_0$,
        it may be interpreted as local radius
of curvature of the magnetic field.

\section{Linear Perturbation}

We now consider a general perturbation of the Vlasov
equation with $f({\bf r, p},t)=f^0({\bf r,p}) +
\delta f({\bf r,p},t)$.  To first order in the
perturbation amplitude $\delta f$ obeys
\begin{eqnarray}
\left({\partial \over \partial t}
+ {\bf v \cdot}{\partial \over \partial {\bf r}} +
{d{\bf p} \over dt}~{\bf \cdot}~
{\partial \over \partial {\bf p}}\right)\delta f
\equiv{D\delta f \over
Dt}=\quad\quad\quad
        \nonumber\\
e(\delta {\bf E} + {\bf v \times} \delta {\bf B})~
{\bf \cdot}~{\partial f^0 \over \partial {\bf p}}~,
\label{exprdf}
\end{eqnarray}
where $\delta{\bf E}$ and $\delta {\bf B}$ are
the perturbations in the electric and magnetic fields.
All scalar perturbation quantities are considered
to have the dependencies
\begin{eqnarray}
F(r)\exp(im\phi +ik_z z -i\omega t)~,
\label{ansatz}
\end{eqnarray}
where the angular frequency $\omega$ is taken to
have at least a small positive imaginary part which corresponds
to a growing perturbation.
          This allows for a correct initial value treatment of
the problem \citep{Landau1946}.
           For a perturbation taken to vanish as $t\to -\infty$,
\begin{equation}
\nonumber
\delta f({\bf r, p},t) =
\end{equation}
\begin{equation}
e\int_{-\infty}^t dt^\prime
\bigg\{\delta {\bf E}[{\bf r}(t^\prime),t^\prime] +
{\bf v}(t^\prime){\times}
\delta {\bf B}[{\bf r}(t^\prime),t^\prime]\bigg\}~{\bf
\cdot}~{\partial f^0 \over \partial {\bf p}}~,
\label{dfLandau}
\end{equation}
where the integration
follows the orbit $[{\bf r}(t^\prime),{\bf p}(t^\prime)]$
which passes through the phase-space point $[{\bf r},{\bf
p}]$ at time $t$.
            For the considered axisymmetric equilibria,
\begin{eqnarray}
{\partial f^0 \over \partial {\bf p}} =
{{\bf p}\over H} \left . {\partial f^0 \over \partial H} \right |_{P_{\phi}}
+ r\hat{\rvecphi~} \left . {\partial f^0 \over \partial P_\phi} \right |_{H}~,
\label{fHP}
\end{eqnarray}
where the partial derivatives are to be evaluated at constant $P_{\phi}$
and $H$, respectively.
Thus, the right-hand side of Eq. (\ref{exprdf}) becomes
\begin{eqnarray}
e\left(
-{d\delta \Phi \over dt}
+i\omega(\dot{\phi}~\delta \Psi-\delta \Phi)
        + i \omega {\bf v}_\perp\cdot\delta {\bf A} \right)
{\partial f^0  \over \partial H} +
        \nonumber\\
e\left( -{d \delta \Psi \over dt}
+im(\dot{\phi}~\delta \Psi -\delta \Phi)
+im{\bf v}_\perp \cdot \delta{\bf A} \right)
{\partial f^0 \over \partial P_\phi}~,
\end{eqnarray}
where $\delta {\bf E}= -{\bf \nabla}\delta \Phi
-\partial {\bf \delta A}/\partial t$ and $\delta {\bf B}
={\nabla \times}{\bf \delta A}$, $\delta \Psi \equiv
r\delta A_\phi$ is the perturbation in the
flux  function, ${\bf v}_\perp =(v_r, v_z)$, and $d/dt =
\partial/\partial t + {\bf v}\cdot {\bf \nabla}$.
         We assume the Lorentz gauge ${\bf \nabla \cdot}\delta
{\bf A} + \partial \delta \Phi/\partial t=0$.

         Evaluating Eq. (\ref{dfLandau}) gives
$$
\!\!\!\delta f=e{\partial f^0 \over \partial H}
\left[-\delta \Phi
+i\omega\int_{-\infty}^t \!\!\!\!\!dt^\prime
\left(\dot\phi^\prime \delta \Psi^\prime
-\delta \Phi^\prime +{\bf
v}^\prime_\perp
\cdot \delta {\bf A}^\prime\right)\right]
$$
\begin{equation}
+e{\partial f^0 \over \partial P_\phi}
\left[-\delta \Psi
+im\int_{-\infty}^t \!\!\!\!dt^\prime~
\left(\dot{\phi}^\prime \delta \Psi^\prime -\delta
\Phi^\prime+{\bf v}^\prime_\perp \cdot \delta
{\bf A}^\prime\right)\right]~,
\label{dfLandaupots}
\end{equation}
where the prime indicates evaluation at
$[{\bf r}(t^\prime),t^\prime]$.
         The integration is along the unperturbed particle orbit so
that $\partial f^0/\partial H$ and $\partial f^0/\partial
P_\phi$ are constants and can be taken outside the
integrals.
Note also that $d/dt$ acting on a function of $({\bf r},t)$
is the same as $D/Dt$.

\section{First Approximation}

As a  starting approximation we neglect (i) the radial
oscillations in the orbits [$(\Delta r/r_0)^2 \ll 1$], (ii) the
self-field corrections to orbits proportional to $\zeta$,  (iii)
the terms in $\delta f$ proportional to $v_\perp^2$ ($v_{th}^2
\approx (\Delta r/r_0)^2 \ll 1$),
(iv) we take $k_z=0$ and (v) we assume the layer is very thin.
         Owing to approximation (iii), we can neglect
the terms $\propto {\bf v}_\perp \cdot \delta {\bf A}$ in
Eq. (\ref{dfLandaupots}) in the evaluation of $\delta \rho$
and $\delta J_\phi$.
          This is because these
terms give contributions to $\delta f$ which
are odd functions of $v_r$ and $v_z$. Therefore, their average
contribution can be neglected. 

         Evaluation of Eq. (\ref{dfLandaupots}) gives
\begin{equation}
\delta f =- e{\partial f^0 \over \partial H}\bigg|_{P_\phi}
\!\!\!
{\dot{\phi}(\omega \delta \Psi-m\delta \Phi) \over
\omega - m\dot{\phi}}
-e {\partial f^0 \over \partial P_\phi}\bigg|_{H}
{\omega \delta \Psi - m \delta \Phi
\over \omega - m \dot{\phi}}~,
\label{dfpot}
\end{equation}
where $\dot{\phi}=\dot{\phi}(r_0)$.
          The approximations lead to a closed system
with potentials $(\delta \Phi,~ \delta \Psi)$
and sources $(\delta \rho,~ \delta J_\phi)$.

We have
$$
\delta \rho =
-e\int d^3p~ \delta f = -{e \over r_0}
\int dp_r dp_zdP_\phi ~ \delta f~,~~~
$$
\begin{equation}
\delta J_\phi =
-e\int d^3p~v_\phi \delta f = -{e \over r_0}
\int dp_r dp_z dP_\phi ~ v_\phi \delta f~.
\label{psints}
\end{equation}
          For the considered distribution function,
Eq. (\ref{equi}), $\partial f^0/\partial H = -f^0/T$.
          The $\partial f^0/\partial P_\phi$ term
in Eq. (\ref{dfpot}) can be integrated by parts.
Furthermore, note that $\partial H / \partial P_\phi =
\dot \phi$ and
$\partial \dot{\phi}/\partial P_\phi =
- (\dot{\phi})^2/H$, which corresponds to an
effective ``negative mass'' for the particle's
azimuthal motion \citep{Kolomenskii1959,Nielson1959,Lawson1988}.
          From the partial integration
the small term proportional to
$\partial v_\phi/\partial P_\phi = v_\phi/(r_0H^3)$
is neglected.
Also note that $H$ is not a constant when performing
the integration over momenta. Evaluating this term by an integration by parts
with a general function $g(P_{\phi})$ in the integrand gives
\begin{eqnarray}
\nonumber
& \ & \int d P_{\phi} \left .
\frac{\partial f^0}{\partial P_{\phi}} \right |_H g(P_{\phi}) =
\\ \nonumber
& - & K \int d P_{\phi} \delta(P_{\phi}-P_0)
\frac{\partial}{\partial P_{\phi}} \left [g(P_{\phi}) e^{-H/T} \right ] =
\\ \nonumber
& - & K \int d P_{\phi} \delta(P_{\phi}-P_0)
\frac{\partial}{\partial P_{\phi}} \left [g(P_{\phi}) \right ] e^{-H/T}
\\
& + & \frac{K}{T} \int d P_{\phi} \delta(P_{\phi}-P_0) g(P_{\phi}) e^{-H/T}
\frac{\partial H}{\partial P_{\phi}}~.
\end{eqnarray}
That is, the integration produces
an additional term which cancels the $1/T$-term.
Thus,
\begin{equation}
\int d P_{\phi} \delta f =
-e \int d P_{\phi} {f^0 \over H}~
{m \dot{\phi}^2(\omega \delta \Psi - m \delta \Phi)
\over (\Delta\omega)^2 }~,
\label{dfpots}
\end{equation}
where $\Delta \omega \equiv \omega -m \dot{\phi}$.
Integrating over the remaining momenta gives
\begin{equation}
(\delta \rho, ~\delta J_\phi)=
(1, v_\phi)~e^2~ n(r)
{m\dot{\phi}^2 \over H} \cdot
{(\omega \delta \Psi -m\delta \Phi ) \over
(\Delta \omega)^2}~.
\label{drhodpot}
\end{equation}

           For a radially thin E-layer we may
take $n(r)=n_0 \exp(-\delta r^2/2\Delta r^2)
\rightarrow n_0\sqrt{2\pi}\Delta r ~\delta(\delta r)$.
   We  comment on this approximation below in more detail
when we include the radial wavenumber $k_r$ of the perturbation.
Then equations (\ref{dpot1}) and (\ref{dpot2}) can be written as
\begin{equation}
[\delta \Phi(r_0),~\delta \Psi(r_0)]=
        \big[1,~r_0 v_\phi (1 + \Delta \tilde \omega) \big]~2\pi^2r_0~ Z
        \int dr ~\delta \rho(r)~,
\label{dpotdrho}
\end{equation}
where $Z \equiv iJ_m(\omega r_0) H_m^{(1)}(\omega r_0)$, 
$\tilde{\omega}\equiv \omega/(m\dot{\phi})$ and
$\Delta \tilde{\omega}\equiv \Delta \omega/(m\dot{\phi})$.
Integrating Eq. (\ref{drhodpot}) over the radial extent of the E-layer
and cancelling out the field amplitudes
gives the dispersion relation
\begin{equation}
1=2\pi^2 r_0~
[n_0 e^2 \sqrt{2\pi} \Delta r]~Z
{m\dot{\phi}^2 \over H} \cdot
{\omega r_0v_\phi (1+\Delta \tilde \omega) -m \over
(\Delta \omega)^2}~.
\label{dispdim}
\end{equation}
In terms of dimensionless variables this becomes
\begin{equation}
1= \pi~ \zeta~ Z~
\left(2 \Delta \tilde{\omega} - {1 \over \gamma^2}\right)
{1 \over (\Delta \tilde{\omega} )^2}~,
\label{disp}
\end{equation}
where $Z=iJ_m(m\tilde{\omega}v_\phi)
H_m^{(1)}(m\tilde{\omega}v_\phi)$,
$H^{(1)}_m=J_m +i Y_m$,
and the field-reversal parameter
$\zeta=4\pi e n_0v_\phi\Delta r\sqrt{\pi/2}/B^e_z$ as given
by Eq. (\ref{zeta1}).

       For $m\gg 1$,
$J_m(m\tilde{\omega}v_\phi)\approx
(2/m)^{1/3}{\rm Ai}(w)$
and $Y_m(m\tilde{\omega}v_\phi)\approx
-(2/m)^{1/3}{\rm Bi}(w)$, where
${\rm Ai}$ and ${\rm Bi}$ are the Airy functions and
$w =(m/2)^{2/3}(\gamma^{-2} -2 \Delta
\tilde{\omega})$ \cite{Abramowitz1965}.
Thus we have $ Z =iJ_mH_m^{(1)} \approx (2/m)^{2/3}[{\rm
Ai}(w){\rm Bi}(w) +i{\rm Ai}^2(w)]$.
It is useful to denote $Z$ as $Z_m(w)$.   For $|w|^2 \gg 1$,
${\rm Ai}(w)\approx (2\sqrt{\pi})^{-1} w^{-1/4}
\exp(-2w^{3/2}/3)$,
${\rm Bi}(w)\approx (\sqrt{\pi})^{-1} w^{-1/4}
\exp(2w^{3/2}/3)$,
$Z_m \approx (2/m)^{2/3}/(2\pi |w|^{1/2})$.

          For $|w|^2 \lesssim 0.5$, ${\rm Ai}(w) = c_1
-c_2 w +{\rm O}(w^3)$ and ${\rm Bi}(w) = \sqrt{3}[c_1+c_2 w
+{\rm O}(w^3)]$, where
$c_1 =1/[3^{2/3}\Gamma(2/3)]\approx 0.355$
and $c_2=1/[3^{1/3}\Gamma(1/3)]\approx 0.259$.
In this limit we have $Z_m(w) \approx (2/m)^{2/3}
[\sqrt{3}(c_1^2-c_2^2 w^2)+i(c_1-c_2 w)^2]$.
          For $|w|^2 \ll 1$, $Z_m \approx (0.347+0.200~i)/m^{2/3}$.

\subsection{Range of Validity}

We are interested in the regime where the wavelength of the
emitted radiation is comparable to the ``bunch length'', i.e.
$\omega \approx m$ or equivalently $\Delta \tilde \omega \ll 1$.
However, Eq. \ref{disp} is only valid if 
$\Delta \tilde \omega \ll \gamma^{-2}$. Since we neglected
$\delta J_r$ and $\delta J_z$ we obtain from the continuity equation
$\delta J_{\phi} = \frac{\omega r_0}{m} \delta \rho$. Due to this
approximation the factor on the right hand side can become bigger 
than the speed of light if $\Delta \tilde \omega > \gamma^{-2}$
which leads to unphysical results. In the latter case 
$\delta J_{\phi} = v_{\phi} \delta \rho$ is a better approximation.
Fortunately, $\Delta \tilde \omega \ll \gamma^{-2}$ is the most
interesting case and in the remainder of this paper we will
always work in this limit.
Furthermore, for the continuum approximation to be valid the mean particle
distance has to be much smaller than the wavelength.

\subsection{Growth Rates}

\def\zm{Z_m}
\def\Zm{{\cal\Zeta}_m}
\def\omtil{\tilde\omega}
\def\domtil{\Delta\omtil}
\def\domtilvert{\vert\domtil\vert}
\def\vth{v_{th}}
\def\sigvert{\vert\sigma\vert}
\def\wvert{\vert w\vert}
\def\Aisq{{\rm Ai^2}(w)}
\def\gvth{\gamma^2\vth^2}
\def\gvthth{\gamma^3\vth^3}

It will prove useful to define two characteristic
values of $m$: $m_1\equiv\zeta^{3/2}\gamma^3$ and
$m_2=2\gamma^3$, and therefore $m_1=\zeta^{3/2}m_2/2$.
We can obtain approximate solutions to Eq. (\ref{disp}) in two
different cases.
There may be solutions with small values
of $\gamma^2\domtil$, so that $w\simeq(m/m_2)^{2/3}$.
In this case, Eq. (\ref{disp}) becomes a simple quadratic
equation, which can be solved for $\domtil$. We can simplify
the solution somewhat by changing variables to $\sigma\equiv\gamma^2
\domtil$ in which case Eq. (\ref{disp}) can be written in the form
\begin{eqnarray}
\nonumber
1={\pi\zeta\zm\gamma^2\over\sigma^2}
(\sigma-1)\approx
-{\pi\zeta\zm\gamma^2\over\sigma^2},
\end{eqnarray}
where we have neglected $\sigma$ compared to one in the
approximate version of this equation.
We find that
\begin{eqnarray}
\sigma\simeq\sqrt{-\pi\zeta\zm\gamma^2}~.
\label{sigsmall}
\end{eqnarray}
For case I let us assume that $m\ll m_2 $, in which case
Eq. (\ref{sigsmall}) implies
\begin{eqnarray}
\nonumber
\sigma\simeq\pm1.121(m_1/m)^{1/3}~e^{i(7\pi/12)}
\\
=1.121(m_1/m)^{1/3}(-0.2588+0.9659i)~.
\end{eqnarray}
so $\sigvert\ll 1$ for $m\gg m_1$. The growth rate of the
unstable mode is
\begin{equation}
\omega_i \simeq {1.083\zeta^{1/2}m^{2/3}\dot\phi\over\gamma}
\end{equation}
in this regime. For case II we assume that $m\gg m_2$, in
which case Eq. (\ref{sigsmall}) implies
\begin{eqnarray}
\nonumber
\sigma=\pm{i\zeta^{1/2}\gamma^{3/2}\over m^{1/2}}=
\pm i\zeta^{1/2}(m_2/2m)^{1/2}
\\
\omega_i \simeq{\zeta^{1/2}m^{1/2}\dot\phi\over\gamma^{1/2}}
~;
\end{eqnarray}
note that the growth rates in cases I and II match almost
exactly at $m=m_2$, where $\sigvert\approx\zeta^{1/2}$.

\begin{figure}
\includegraphics[width=3.3in]{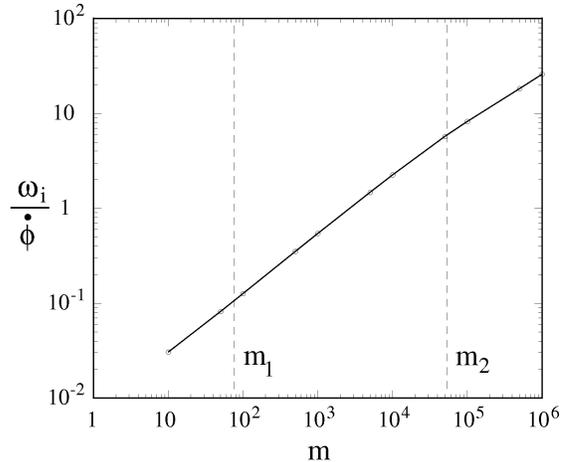}
\caption{The graph shows
the frequency dependence of the growth
rate
for a sample case where $\gamma=30$ and
$\zeta =0.02$ obtained from our approximations
for Eq. (\ref{disp}). For these parameters,
$m_1\approx 10^2$ and $m_2 \approx 2.7 \times 10^4$.
}
\label{growthnobeta}
\end{figure}

Note that $m_2 \dot{\phi}$ is the approximate frequency
of the peak of the single particle synchrotron
radiation spectrum.
For more accurate results we employ a numerical method for
solving Eq. \ref{disp} outlined in \cite{Botten1983}.
This method also allows us to count the
number of roots which are enclosed
by a contour. So far we have no
numerical evidence of the
existence of more than one
solution with a positive real part.
The numerical results agree very well with our approximations
even if $m < m_1$ and are shown in Fig. \ref{growthnobeta}.

\subsection{Comparison with Goldreich and Keeley}

       Goldreich and Keeley
\cite{GoldreichKeeley1971} find a
radiation instability in
a thin ring of relativistic,
monoenergetic, zero temperature
electrons constrained to
move in a circle of fixed radius. Under
the condition $1 \ll m^{1/3} \ll
\gamma$ their growth rate is
$\omega_i \approx 1.16\dot{\phi}~
m^{2/3} [r_eN/(\gamma^3r_0)]^{1/2}$
which is close to our growth
rate with $L$ replaced by $r_0$.

%%%%%%%%%%%%%%%%%%%%%%%%%%%%%%%%%%%%%%%%%%
%%%%%%%%%%%%%%%%%%%%%%%%%%%%%%%%%%%%%%%%%

\section{Nonlinear Saturation}

           Clearly the rapid exponential growth of the
linear perturbation can continue only for a
finite time.
          We analyze this by studying
the trapping of electrons in the moving
potential wells of the perturbation.
For $(\Delta r/r_0)^2 \ll 1$, the electron orbits
can be treated as circular.
           The equation of motion is
\begin{eqnarray}
{d P_\phi \over dt} = r \delta F_\phi~,\quad
\delta F_\phi = -e[\delta E_\phi +
({\bf v \times }\delta {\bf B})_\phi]~,
\label{eom}
\end{eqnarray}
where $P_\phi$ is the canonical angular momentum, where
\begin{eqnarray}
\delta F_\phi=
-e \delta E_{\phi 0} \exp(\omega_i t)
\cos(m\phi -\omega_r t) ~,\!\!
\end{eqnarray}
where $\delta E_{\phi 0}$ is the initial value of
the potential,
        $\omega_r \equiv {\rm Re}(\omega)$, and
$\omega_i \equiv {\rm Im}(\omega)$.

For a relativistic particle in a circular orbit,
\begin{eqnarray}
\delta P_\phi = m_{e*} r_0^2 \delta \dot{\phi},~~
{\rm where}~ m_{e*}={- m_e \gamma^3 \over \gamma^2-1}
\approx - m_e \gamma,
\label{negmass}
\end{eqnarray}
where $m_{e*}$ is the ``effective mass,'' which
is negative, for the azimuthal motion of the electron
(\cite{Kolomenskii1959,Nielson1959} or \cite{Lawson1988}, p.68).
Combining equations (\ref{eom}) and (\ref{negmass}) gives
\begin{equation}
{d^2 \varphi \over dt^2}= -\omega_T^2(t)
\sin \varphi~,
\end{equation}
where $
\varphi \equiv m\phi -\omega_t t + \frac{3}{2} \pi$,
$ \omega_T\equiv \omega_{T0}\exp(\omega_i t/2),$ and $
\omega_{T0} \equiv [ e m \delta E_{\phi 0} / ( m_e \gamma
r_0) ]^{1/2}$,
where $\omega_T$ is termed the ``trapping frequency.''
At the ``bottom'' of the potential well of the
wave, $\sin\varphi \approx \varphi$.
An electron  oscillates about the bottom of the
well with an
angular frequency $\sim \omega_T$.
          This is of course a nonlinear effect of the
finite wave amplitude.
          A WKBJ solution of Eq. (\ref{negmass})
gives
\begin{equation}
\varphi \propto\omega_{T0}^{-1/2} \exp(-\omega_i t/4)
\sin\big\{(2\omega_{T0}/\omega_i)[\exp(\omega_i t/2)
-1]\big\}~.
\end{equation}
         The exponential growth
of the linear perturbation will cease at the time
$t_{sat}$ when the particle is turned around in the
potential well.
         This condition corresponds
to $\omega_T(t_{sat}) \approx \omega_i$.
         Thus, the saturation amplitude is
\begin{eqnarray}
\big|\delta E_{sat}\big|^2 =
\left ( \frac{m_e\gamma }{ e r_0 m } \right )^2
\left({\omega_i(m) \over \dot{\phi}}\right)^4~,
\label{satamp}
\end{eqnarray}
where $|\delta E_{sat}|\equiv
|\delta E(t_{sat})|=|\delta E_0|\exp(\omega_i
t_{sat})$.

\section{First Approximation with $k_z \neq 0$}

Here, we consider $k_z \neq 0$ but keep the
other approximations. Our ansatz for $\delta f$
is general enough to handle this case since it
retains the biggest contribution to the Lorentz force in
the $z$-direction which is of the order $v_{\phi} B_r$.
In place of Eq. (\ref{dfpots}) we obtain
\begin{equation}
\int d P_{\phi} \delta f =
-e \int d P_{\phi} {f^0 \over H}~
{m \dot{\phi}^2(\omega \delta \Psi - m \delta \Phi)
\over (\omega-m\dot{\phi} - k_z v_z)^2 }~,
\end{equation}
where we assume without loss of generality $k_z>0$
and $k_z \ll m/r_0, \omega$.
         In place of Eq. (\ref{disp}) we find
\begin{equation}
\varepsilon(\omega,k_z)=1\! +
k_zA(\omega,k_z)\int_{-\infty}^\infty \!\!dv_z
~{\exp(-v_z^2/2v_{th}^2)
\over \sqrt{2\pi}~ v_{th}}\big[...\big]
=0,
\end{equation}
where
$$
\big[...\big] \equiv
-{m \dot{\phi} \over (\omega-m\dot{\phi}-k_zv_z)^2}~.
$$
Here, $\varepsilon$ acts as an effective dielectric
constant for the E-layer,  and
$$
A(\omega,k_z) \equiv \pi ~\zeta ~Z(\omega,k_z)~
\left(u-{k_\phi \over k_z \gamma^2}\right),
~ ~ u\equiv
{\omega - m\dot{\phi} \over k_z}~,
$$
\begin{equation}
Z \equiv iJ_m[r_0(\omega^2-k_z^2)^{1/2}]~
H_m^{(1)}[r_0(\omega^2-k_z^2)^{1/2}]~,
\end{equation}
and $k_\phi = m/r_0$ is the azimuthal wavenumber.
The expression for $Z$ is from \S 4.
          An integration by parts gives
\begin{eqnarray}
\varepsilon(u)=1 +
A(\omega)\int dv_z~{\exp(-v_z^2/2v_{th}^2)
\over \sqrt{2\pi}~ v_{th}^3}
{m\dot{\phi} v_z/k_z \over v_z- u} ~,
\end{eqnarray}
where the $k_z$ dependence of $\varepsilon$
and $A$ is henceforth implicit.
          We can also write  this equation as
\begin{eqnarray}
\varepsilon(u)=1 + B(u) \left [ 1 +
\frac{u}{v_{th}}
F\left({u\over v_{th}}\right) \right ]~,
\label{epsu}
\end{eqnarray}
where
\begin{eqnarray}
\nonumber\\
B(u) &\equiv& {\pi \over v_{th}^2}~\zeta~Z~{k_\phi \over k_z}
\left(u- {k_\phi \over k_z \gamma^2}\right)~,
\end{eqnarray}
and
$$
         F(z) \equiv
{1 \over \sqrt{2\pi}}\int_{-\infty}^\infty
dx ~{\exp(-{x^2 / 2}) \over x-z}~,
$$
for ${\rm Im}(z)>0$, and
$$
         F(z) \equiv
{1 \over \sqrt{2\pi}}\int_{-\infty}^\infty
dx ~{\exp(-{x^2 / 2}) \over x-z}+i\sqrt{2\pi}
\exp\left(-{z^2 \over 2}\right)~,
$$
${\rm for~Im}(z)<0$.
        The second expression for $F(z)$ is the
analytic continuation of the first expression
to ${\rm Im}(z)<0$ which corresponds to wave
damping
(see, e.g., \cite{Montgomery1964}, ch. 5).
Note that terms of order $\Delta \tilde \omega$
have been omitted.

          For $m\gg1$, the factor $Z=iJ_m(J_m+iY_m)$ can
be expressed in terms of Airy functions in a way
similar to that done in \S 5.
           One finds
$J_m[r_0(\omega^2-k_z^2)^{1/2}] \approx
(2/m)^{1/3}{\rm Ai}(w)$,
$Y_m[r_0(\omega^2-k_z^2)^{1/2}] \approx
-(2/m)^{1/3}{\rm Bi}(w)$,
\begin{equation}
        Z_r\approx \left({2\over m}\right)^{2/3}{\rm Ai}(w){\rm
Bi}(w)~,~ Z_i\approx \left({2\over
m}\right)^{2/3}{\rm Ai}^2(w)~,
\label{ZriAiBi}
\end{equation}
        where
$$
        \tan\psi \equiv {k_z \over k_\phi}~,~
w \equiv \left({m \over 2}\right)^{2/3}
\left({1\over \gamma^2} +\tan^2\psi
-2u\tan\psi\right)~.
$$
It is clear that $\varepsilon$ has in general a rather
complicated dependence on $u=u_r+iu_i$ and $\tan \psi$.
          Note that the expression for $w$ goes
over to our earlier $w$ for $\psi=0$
noting that $u\tan \psi \rightarrow \Delta
\tilde{\omega}$.

%%%%%%%%%%%%%%%%%%%%%%%%%%%%%%%%%%%%%%%%%%
\begin{figure}
\includegraphics[width=3.3in]{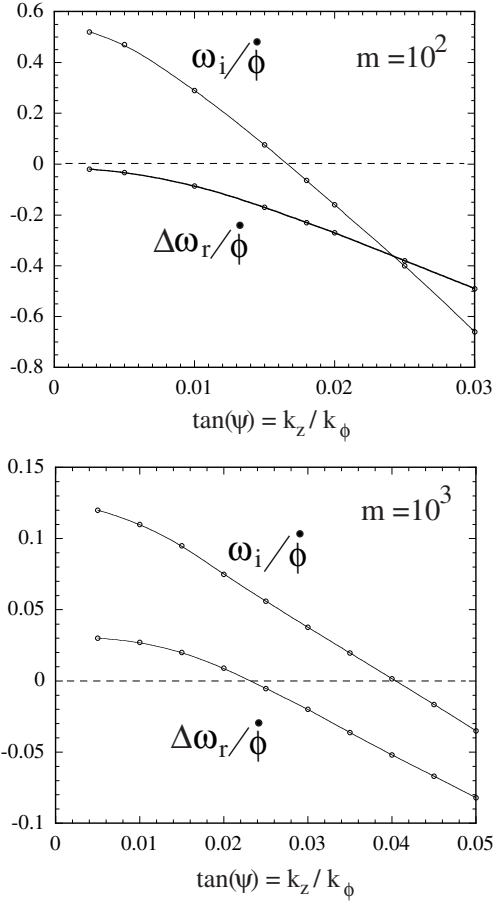}
\caption{The figure shows
the growth/damping rate $ \omega_{i}$
and real part of the
frequency $\Delta \omega_{r}=\omega-m\dot{\phi}$
in units of $\dot{\phi}$ as
a function of $\tan\psi=k_z/k_\phi$ for $m=100$
and $m=1000$ for an E-layers
with $\gamma=30$, $\zeta=0.02$ and $v_{th}=30/\gamma^2$.
In the region of damping $\omega_i<0$, the
second expression for $F(z)$ in Eq. (\ref{epsu}) is used.}
\label{growthkz}
\end{figure}
%%%%%%%%%%%%%%%%%%%%%%%%%%%%%%%%%%%%%%%%%

          A limit where Eq. (\ref{epsu}) can be solved
analytically is for $|u|^2= |\Delta \tilde{\omega}|^2/
\tan^2\psi \gg v_{th}^2$, that is, for
sufficiently small $\tan \psi$.
          In this limit Eq. (\ref{epsu}) can be expanded
as an asymptotic
series $F(z)=-1/z - 1/z^3 - 3/z^5 -..$.
          Keeping
just the first three terms of the expansion gives
\begin{equation}
\varepsilon =1 + \pi \zeta Z
\left(1+{3 v_{th}^2 \tan^2\psi \over (\Delta \tilde{\omega})^2}\right)
{{\gamma^{-2}}\over(\Delta \tilde{\omega} )^2} = 0.
\label{dispkzapprx}
\end{equation}
         For $\tan\psi \rightarrow 0$ and $\Delta \tilde \omega \ll \gamma^{-2}$, this is
the same as Eq. (\ref{disp}) as it should be.
In general Eq. \ref{dispkzapprx} will have more than one unstable mode.
In the remainder of this paragraph we will only study the largest unstable solution
for which we recover the growth rates found in \S 4 in the limit $\tan\psi \rightarrow 0$.
Fig. \ref{growthkz} shows some sample solutions.
        For the case
shown the $u$ dependence of $Z$ is negligible.

         General solutions of Eq. (\ref{epsu}) can be obtained
using the Newton-Raphson method (\cite{Teukolsky1989},
ch. 9) where an
initial guess of $(u_r,u_i)$ gives $(\epsilon_r,\epsilon_i)$.
This guess is incremented by an amount
\begin{eqnarray}
          \left[\begin{array}{c} {\delta u_r}
\\ \noalign{\medskip}{\delta u_i} \end{array}\right]
~= ~\left [\begin {array}{cc}
{\partial \epsilon_r/\partial u_r}&{\partial
\epsilon_r/\partial u_i}
\\\noalign{\medskip}{\partial \epsilon_i/\partial u_r}&
{\partial \epsilon_i/\partial u_i}
\end {array}\right ]^{-1}
\left[\begin{array}{c} {-\epsilon_r}
\\ \noalign{\medskip}{-\epsilon_i} \end{array}\right]~,
\end{eqnarray}
and the process is repeated until $\varepsilon_r=0$
and $\varepsilon_i=0$.
         Fortunately, the convergence
is very rapid and gives
$|\varepsilon| < 10^{-10}$ after a few iterations.

\begin{figure}
\includegraphics[width=3.2in]{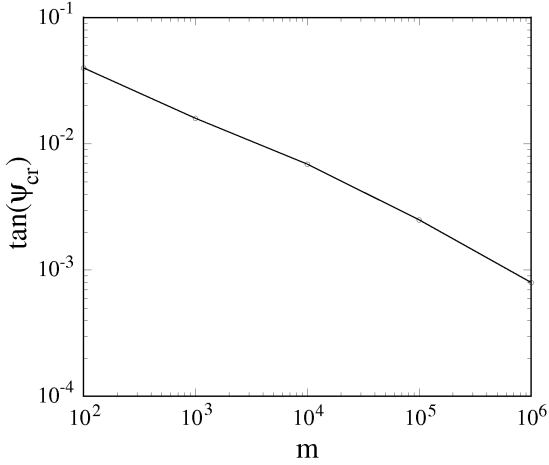}
\caption{Critical angle for
$\gamma=30$, $\zeta=0.02$ and $v_{th}=30/\gamma^2$.}
\label{critang}
\end{figure}

         Fig. \ref{growthkz} shows the dependence of the
complex wave frequency on the tangent
of the propagation angle,
$\tan \psi = k_z/k_\phi$, for a sample cases.
          The maximum growth rate is for $\psi =0$
or $k_z=0$.
          With increasing $\psi$ the growth rate
decreases, and for $\psi$ larger
than a critical angle $\psi_{cr}$ there is
damping.  For the damping the second expression
for $F$ in Eq. (\ref{epsu}) must be used.
          Roughly, we find that the critical angle
corresponds to having the wave phase velocity
in the $z-$direction of the order of the thermal
spread in this direction,  that is,
$u_r = \Delta \omega_r/k_z \sim v_{th}$.
This gives
\begin{eqnarray}
\tan\psi_{cr} \sim {\sqrt{\zeta}
\over v_{th} \! ~ \! \gamma ~ m^{1/3}}=
\left({ r_e N \over v_{th}^2
\gamma^3 L}\right)^{1/2} \!
{1\over m^{1/3}} \le \frac{1}{\gamma^2 v_{th}}~,
\label{tanpsi}
\end{eqnarray}
for $m_1<m<m_2$. Note that the dimensionless parameter
which determines the cut-off at $\tan \psi_{cr}$ is $\gamma^2 v_{th}$.
Our numerical calculations of
$\psi_{cr}$ give a slightly
faster dependence, $\tan \psi_{cr}
\propto 1/m^{0.40}$
for this range of $m$. Fig. \ref{critang} shows
the $m$-dependence of the critical angle.
It is reasonable to assume that in a particle accelerator
the weak focusing in the z-direction sets a low limit on $k_z$.

%%%%%%%%%%%%%%%%%%%%%%%%%%%%%%%%%%%%%%%%%%
%%%%%%%%%%%%%%%%%%%%%%%%%%%%%%%%%%%%%%%%%

\subsection{Nonlinear Saturation for $k_z \neq 0$}

We generalize the results of \S 6 by
including the axial as well as the azimuthal
motion of the electrons in the wave.
          The axial equation of motion is
\begin{eqnarray}
m_e\gamma {d^2 z \over dt^2}
&=&-e\left[\delta E_z +({\bf v}
\times \delta {\bf B})_z\right]
\nonumber\\
& \approx & -e \delta E_{z 0}
\exp (\omega_i t) \cos(m\phi+k_z z -\omega t)~.
\label{axialeom1}
\end{eqnarray}
         The approximation involves neglecting the force
$\propto v_r \delta B_\phi$ which is valid for
a radially thin layer ($\Delta r^2/r_0^2 \ll 1$).
           Following the development of \S 6,
the azimuthal equation of motion
is
\begin{eqnarray}
m_e\gamma r_0{d^2 \phi \over dt^2}
=-e \delta E_{z 0} \cos(m\phi+k_z z -\omega t)~.
\label{axialeom2}
\end{eqnarray}
Combining equations (\ref{axialeom1})
and (\ref{axialeom2}) gives
\begin{eqnarray}
{d^2\varphi \over dt^2}=-~{e~ m \delta E_{\phi 0} \over
m_e \gamma r_0}\left(1+\tan^2\psi\right)
\sin\varphi~,
\end{eqnarray}
where $\varphi\equiv m\phi+k_zz-\omega t +\frac{3}{2} \pi$
and $\tan\psi = k_z/k_\phi$.
          Because $\psi^2 \ll 1$ for wave growth
(Eq. (\ref{tanpsi})), the saturation
wave amplitude $\delta E_{sat}$
is again given by Eq. (\ref{satamp}).

\section{Thick Layers Including Radial Betatron Oscillations}
\subsection{The Limit $k_r \Delta r \gg 1$}

In this section we include the
small but finite radial thickness of the E-layer.
         We keep the other approximations mentioned
at the beginning of \S 5.
        In particular we consider
$k_z=0$.
         In order to include the layer's radial thickness, we
consider the wave equations within the E-layer,
\begin{eqnarray}
(\nabla^2 + \omega^2)\delta \Phi
&=& -4\pi \delta \rho~,
\nonumber \\
(\tilde{\nabla}^2 +\omega^2)\delta \Psi
&=& -4\pi r \delta J_\phi~,
\label{laplacians}
\end{eqnarray}
where
\begin{equation}
\tilde{\nabla}^2 \equiv {\partial^2 \over \partial r^2}
-{1 \over r}{\partial \over \partial r} - \frac{m^2}{r^2}
+{\partial^2 \over \partial z^2}~,
\end{equation}
is the adjoint Laplacian operator.

        Within the E-layer, we assume
that the potentials can be written
in a WKBJ expansion as
\begin{equation}
(\delta \Phi,~\delta \Psi) =
(K_\Phi,~K_\Psi)\exp\big[im\phi
+ik_r (r-r_0) -i\omega t\big]~,
\label{WKBJ}
\end{equation}
where $k_r$ is the radial wavenumber with
$(k_r \Delta r)^2 \gg 1$
$(K_\Phi,K_\Psi)$ are constants.
        This is equivalent to assuming
that the charge density
is constant between $r_0-\Delta r$
and $r_0+\Delta r$ and zero elsewhere.
Evaluation of the time integrals in
Eq. (\ref{dfLandaupots}) for $r=r_0$  gives
\begin{eqnarray}
\int_{-\infty}^t dt^\prime \delta \Phi^\prime =
\delta \Phi(r_0,t)\times \quad\quad
\quad
\nonumber\\
\sum_{n=-\infty}^\infty {J_n(k \delta r_i)i^n
\exp(-ik_\phi \delta r_i -in\psi)
\over i(m \dot{\phi} + n \omega_{\beta r} -\omega)}~,
\label{Psiint}
\end{eqnarray}
where $n$ is an integer, $k\equiv (k_r^2+k_\phi^2)^{1/2}$,
with $k_\phi = m/r_0$, and $\tan \psi \equiv k_r/k_\phi$.
       There is an analogous expression for the integral
of $\delta \Psi$.
          We have used Eq. (\ref{rorbit})
for the radial motion with $\varphi =0$
assuming $\zeta^2 \ll 1$ and
$\zeta \ll \gamma^2(\Delta r/r_0)$ so
that $\omega_{\beta r} =1/r_0$, and Eq.
(\ref{phiorbit}) for the $\phi$-motion with
       $ \partial \dot{\phi}_0/\partial r|_{r_0}/\omega_{\beta r}
= - 1/r_0$.
Using equations
(\ref{dfLandaupots}) and (\ref{Psiint}), the momentum space
integrals (\ref{psints}) can be done to give
\begin{widetext}
\begin{eqnarray}
\nonumber
\left ( eK e^{-H/T} \right )^{-1} \int  d P_{\phi} \delta f =
\frac{1}{T} (K_\Psi\dot \phi -K_\Phi)\left \{
J_0(k \delta r_i) - 1 + (m \dot \phi - \omega) \sum_{n=-\infty}^{~~\infty~~\prime}
\frac{i^n e^{-in \psi -i k_{\phi} \delta r_i} J_n(k \delta r_i)}
{m \dot \phi + n \omega_{\beta r} - \omega}  \right \} -
\\ 
\frac{m \dot \phi^2 }{H} \frac{\omega K_{\Psi}-m K_{\Phi}}{(m \dot \phi - \omega)^2} -
\frac{m \dot \phi^2 }{H} m( K_{\Psi} \dot \phi - K_{\Phi} ) \left \{
\frac{J_0(k \delta r_i) - 1}{(m \dot \phi - \omega)^2} +
\sum_{n=-\infty}^{~~\infty~~\prime} \frac{i^n e^{-in \psi -i k_{\phi} \delta r_i}
J_n(k \delta r_i)}{(m \dot \phi + n \omega_{\beta r} - \omega )^2} \right\}
~,
\end{eqnarray}
and finally if $\Delta \tilde \omega \ll \gamma^{-2}$
\begin{eqnarray}
\delta \rho \approx {e^2 n_0  m \dot{\phi}^2K_\Phi \over H}
\bigg( { r_0^2  \dot{\phi} \omega
-m[1-(1-F_0)/\gamma^2]\over
(\omega - m\dot{\phi})^2}  -  {m\over \gamma^2}
\sum_{n=-\infty}^{~~\infty~~\prime}
{F_n  \over (m\dot{\phi}+n \omega_{\beta r} -\omega)^2} \bigg)~.
\label{drhotthick}
\end{eqnarray}
\end{widetext}
The prime on the sums indicate
that the $n=0$ term is omitted.
Here,
\begin{equation}
F_n\equiv {i^n \exp(-in\psi)\over \sqrt{2\pi} \chi}
\int_{-\infty}^\infty d\xi~
J_n(\xi) \exp\left(- {\xi^2 \over 2 \chi^2}
-i{k_\phi  \xi \over k} \right)~,
       \end{equation}
with
\begin{equation}
\chi \equiv k\Delta r ~.
\end{equation}
The $1/T$ terms in Eq. \ref{drhotthick} do not cancel exactly.
      They may  be neglected if
\begin{equation}
|\Delta \tilde \omega|^2 \ll {F_0}v_{th}^2
\end{equation}
for the $n=0$ term or if
\begin{equation}
{|\Delta \tilde \omega|}|n/m - \Delta \tilde \omega| \ll {v_{th}^2}
\end{equation}
for the $n \neq 0$ terms.
      
For  weak E-layers
we have for
$\chi \rightarrow 0$,  $F_0 \rightarrow 1$
and $F_{n\neq 0} \rightarrow 0$.
     In this limit we
recover the results of \S 4.
For $\chi \gg 1$ and
$1 \ll  k_r r_0 \ll  k_{\phi} r_0$, the Gaussian
factor in the integrand of $F_n$
   can be neglected so that one obtains
$$
F_n \approx
i^n e^{-in \psi} \frac{1}{\sqrt{2 \pi} \chi}
\frac{2k}{|k_r|} \cos
\left ( \frac{n \pi}{2} \right )~,~~
{\rm even}~  n~,
$$
\begin{equation}
F_n \approx- i^n e^{-in \psi}
\frac{1}{\sqrt{2 \pi} \chi} \frac{2ik}{|k_r|} \sin
\left ( \frac{n \pi}{2} \right )~,~~
{\rm odd} \ n~.
\end{equation}

\begin{figure}
\includegraphics[width=3.2in]{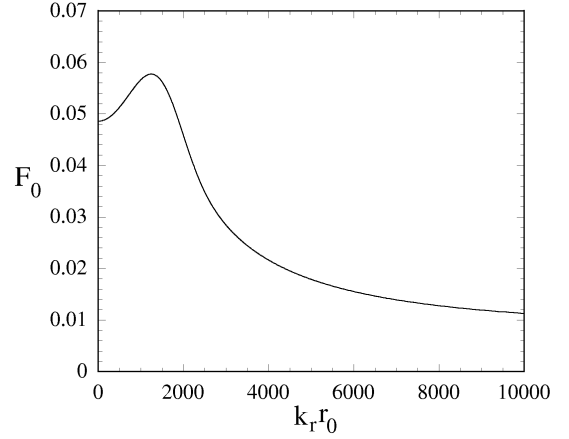}
\caption{$F_0$ for $v_{th}=0.01$ and $ k_{\phi}r_0=10^4$}
\label{figF0}
\end{figure}

     An alternative approximation
for $F_n$ can be obtained
by using the
   integral representation of
the Bessel function.
The remaining integral can then
be computed numerically more easily.
In this way we find
    \begin{equation}
F_n = \frac{i^n e^{-in \psi}}{2\pi} \int _{-\pi }^\pi
d\theta \exp\left[-in \theta -(\chi^2/2)  ( k_{\phi}/k - \sin
\theta  )^2\right]~.
\label{Fnnum}
    \end{equation}
For $\chi \gg 1$, $1 \ll (k_{\phi}/k_r)^2$ and $|n| < \sqrt{\chi}$
we can approximate $\sin \theta$ in the exponent by a parabola
at its maximum. We obtain
     \begin{equation}
F_n \approx \frac{i^n e^{-in \psi}}{2^{3/4}
\Gamma\left(\frac{3}{4}
\right ) \sqrt{\chi}}~.
\end{equation}
     In general $ F_n/(i^n e^{-in \psi})$
decreases as $\chi$ and $n$ increase.
     This acts to prevent the
unlimited increase of
the growth rate as $m \longrightarrow \infty$,
and it ensures that the sums over $n$ converge.
Fig. \ref{figF0} shows a plot of $F_0$ obtained
by numerical evaluation of
   Eq. \ref{Fnnum}.
%%%%%%%%%%%%%%%%%%%%%%%%%%%%%%%%%%%%%%%%%%
%%%%%%%%%%%%%%%%%%%%%%%%%%%%%%%%%%%%%%%%%%

Within the E-layer, Eq. (\ref{laplacians}) gives
\begin{eqnarray}
k_r^2=\omega^2-{m^2 \over r_0^2}\quad\quad\quad
\quad\quad\quad\quad\quad\quad\quad\quad\quad \quad
\nonumber \\
+{4\pi e^2 n_0  m \dot{\phi}^2
\over H}
\bigg({r_0^2\dot{\phi}\omega -m[1-(1-F_0)/\gamma^2] \over
(\omega - m\dot{\phi})^2}
\nonumber\\
-~
{m\over \gamma^2}{\sum_n}^\prime {F_n \over
(m\dot{\phi}+n\omega_{\beta r} -\omega)^2} \bigg)~.
\label{calckr}
\end{eqnarray}
In terms of dimensionless variables this
equation becomes
\begin{eqnarray}
\nonumber
{\bar k_r}^2 = 2m^2 \Delta
\tilde \omega -\frac{m^2}{\gamma^2} +\quad\quad\quad
\quad\quad\quad\quad\quad\quad\quad\quad
\\ \nonumber
\frac{\zeta \sqrt{2}}{v_{\phi}
v_{th} \sqrt{\pi}} \left (
\frac{(1+ \Delta \tilde \omega)
(1 - \gamma^{-2})
- [1+(F_0 - 1)/ \gamma^2]}{(\Delta \tilde
\omega)^2}
\right . \\ \left .
- \frac{1}{\gamma^2} {\sum_n} '
\frac{F_n}{( v_{\phi}^{-1} n/m -
\Delta \tilde \omega)^2} \right )~,\quad~~
\label{calckrbar}
\end{eqnarray}
where $\bar k_r \equiv r_0 k_r$,
$ \bar k_{\phi} \equiv r_0 k_{\phi}$,
$\bar k \equiv r_0 k$, and $\chi = \bar k v_{th}$.

         Notice that
Eq. \ref{WKBJ} can also be written
as
\begin{equation}
\delta \Phi = C_2\sin\big[k_r(r-r_0)\big]
+C_3\cos\big[k_r(r-r_0)\big]~,
\end{equation}
for $r_0-\Delta r \leq r \leq r_0+\Delta r$.
         For $r\leq r_0-\Delta r$, we have
\begin{equation}
\delta \Phi = C_1 J_m(\omega r)~,
\end{equation}
since the potential must be well behaved
as $r \rightarrow 0$.
         For $ r\geq r_0+\Delta r$, we must have
\begin{equation}
\delta \Phi = C_4 \big[J_m(\omega r) + i Y_m(\omega r)\big]~.
\end{equation}
This combination of Bessel functions gives
$\delta \Phi(r\rightarrow \infty) \rightarrow 0$
for the assumed conditions where ${\rm Im}(\omega)>0$.
Note that these potentials are just the solutions
of Eq. (\ref{laplacians}) in our approximation for
$\delta \rho$. The eigenvalue problem can now be solved
by matching the boundary conditions. However,
we have not solved the full eigenvalue problem.
        Instead we consider unstable solutions
with the restriction that
$k_r \Delta r \gg 1$.
        Under this condition
we can interpret Eq. \ref{calckrbar} as a local
dispersion relation.  Unstable modes found
from Eq. \ref{calckrbar} will need a slight
correction in order to satisfy the boundary conditions.

We expect that Eq. \ref{calckrbar} has solutions near
each betatron resonance at $\Delta \tilde \omega = {\pm n}/{m}$.
This is a familiar concept in the treatment of resonances in storage
rings (cf. \cite{chao} or \cite{schmekel}).
We extract each solution by summing over a single value
of $n$ and $-n$ only and obtain from Eq. \ref{calckrbar}
for the case $n \neq 0$ and $\Delta \tilde \omega \ll \gamma^{-2}$
\begin{eqnarray}
\nonumber
\Xi \equiv - \frac{\gamma^2 v_{th} \sqrt{\pi} 
\left ( \bar k_r^2 +m^2 \gamma^{-2}\right )}{\zeta \sqrt{2}} =
\\ 
\frac{F_{-n}}{\left ( \frac{n}{m} + \Delta \tilde \omega \right )^2} +
\frac{F_{n}}{\left ( \frac{n}{m} - \Delta \tilde \omega \right )^2} ~.
\end{eqnarray}
Thus,
\begin{eqnarray}
\Delta \tilde \omega \approx \frac{F_{-n}-F_n}{\Xi} \pm \frac{n}{m}
\end{eqnarray}
for sufficiently big $\Xi$, i.e. we expect the imaginary part of
$\Delta \tilde \omega$ to be negligible for the $n \neq 0$ modes.
Despite a lot of effort we were not able to prove this statement
under more relaxed conditions.

We can easily find an analytic
solution of Eq. \ref{calckrbar}
for the case where the $n=0$ term is dominant.
If $|\Delta \tilde \omega| \ll 1 / \gamma^2$ and
$|\Delta \tilde \omega| \ll F_0 / \gamma^2$, we obtain
\begin{eqnarray}
\Delta \tilde \omega = \pm \frac{2^{1/4} \sqrt{-\zeta F_0}
}{\pi^{1/4} \sqrt{v_{th} (m^2+\gamma^2 \bar k_r^2)}}
\label{soln0}
\end{eqnarray}
The dependence of the growth rate on $k_r$ becomes
significant when
$\gamma^2 \bar k_r^2 / m^2$ is comparable to unity.
For $m\sim m_2\sim \gamma^3$, we see that this happens
when $(k_r\Delta r)^2/\gamma^4v_{th}^2\sim 1$,
which involves the combination $\gamma^2 v_{th}$ again.

\begin{figure}
\includegraphics[width=3.5in]{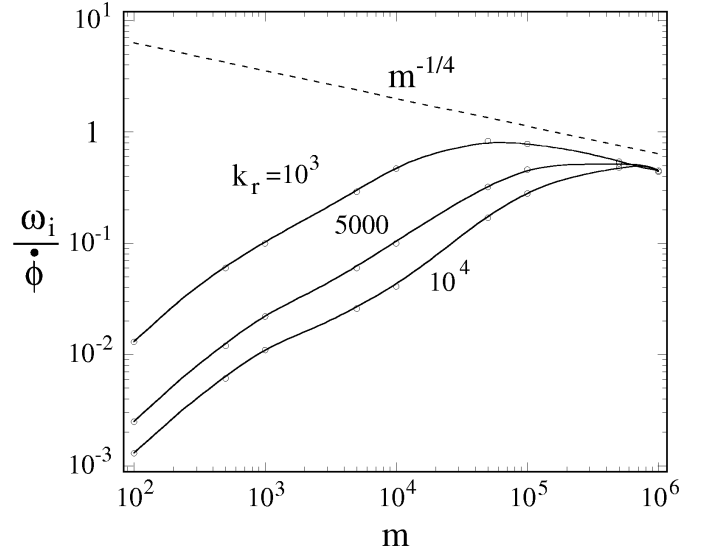}
\caption{Growth rates in the limit $\bar k_r v_{th} \gg 1$
for our reference case $\gamma=30$, $\zeta=0.02$ and
$v_{th}=1/\gamma^2$ and various values of $\bar k_r$. The line
proportional to $m^{-1/4}$ is shown for comparison.}
\label{krbvthgg1}
\end{figure}

The growth rate of Eq. \ref{soln0} is proportional to $\sqrt{\zeta}$.
     This  implies from \S 5 that
the emitted power scales as the square of
the number of particles in the E-layer
which corresponds to coherent radiation.
       Sample results are
shown in Fig. \ref{krbvthgg1}.
      We conclude that
the main effect of the betatron oscillations
is an indirect one.
    The radial motion itself is
unimportant for the interaction.
    However, the influence of the radial motion
on the time dependence of
the azimuthal angle
$\phi$ of a particle is important since a
shift in $\phi$ can take the particle  out
of coherence with the wave.
      This effect is accounted for
by $F_0$.

\subsection{Qualitative Analysis of the Effect of the Betatron Motion}

Let us suppose that $v_{th} \gg 1/\gamma^2$, and that
$|\Delta \tilde \omega|$ is not necessarily small compared with $v_{th}$
(We can still assume $|\Delta \tilde \omega| \ll 1$ without requiring the
more restrictive condition $\gamma^2 | \Delta \tilde \omega| \ll 1$.).
The key effect of the betatron oscillations is to ``wash out'' the 
phase coherence
of the response within the layer; for a cold layer, all orbiting particles
move in ``lock step'', which is particularly favorable for a bunching 
instability.
Let us suppose that $|\Delta \tilde \omega|$ has a real part that is
substantially larger than $1/\gamma^2$. The response in the layer scales as
an Airy function with argument $w(1+\xi)$ where $|\xi| < v_{th}$.
The phase accumulated across the layer thickness $\sim v_{th} r_0$ is
$\eta \sim m v_{th}^{3/2}$ if $\Delta \tilde \omega_r \ll v_{th}$
and $\eta \sim m v_{th}^{3/2} (\Delta \tilde \omega_r / v_{th})^{1/2}$
if  $\Delta \tilde \omega_r \gg v_{th}$.
Large $\eta$ ought to imply substantial decoherence of the response 
in the layer.
We see that this is likely irrespective of the value of
$\Delta \tilde \omega_r / v_{th}$ provided that $m \gg v_{th}^{-3/2}$, i.e. for
$m/\gamma^3 \gg (\gamma^2 v_{th})^{-3/2}$. At large values of 
$\gamma^2 v_{th}$, phase
smearing should suffice to suppress - if not eliminate - the bunching 
instability at
frequencies near the synchrotron peak. Moreover, if $\gamma^2 v_{th} 
\gtrsim \zeta^{-1}$,
the instability should be suppressed over the entire range $m \gtrsim 
\zeta^{3/2} \gamma^3$
for which we found unstable modes in \S 4. Large $\Delta \tilde 
\omega_r / v_{th}$
would merely accentuate the smearing. At a given value of $m$, we see that
$\Delta \tilde \omega_r \gtrsim (m^2 v_{th}^2)^{-1}$, i.e.
$\gamma^2 \Delta \tilde \omega_r \gtrsim (m/\gamma^3)^{-1}(\gamma^2 
v_{th})^{-2}$
suffices for large phase decoherence in the layer.

%%%%%%%%%%%%%%%%%%%%%%%%%%%%%%%%%%%%%%%%%%%%%%
%%%%%%%%%%%%%%%%%%%%%%%%%%%%%%%%%%%%%%%%%%%%%

\subsection{The Limit $k_r \Delta r \ll 1$}
\begin{figure}
\includegraphics[width=3.5in]{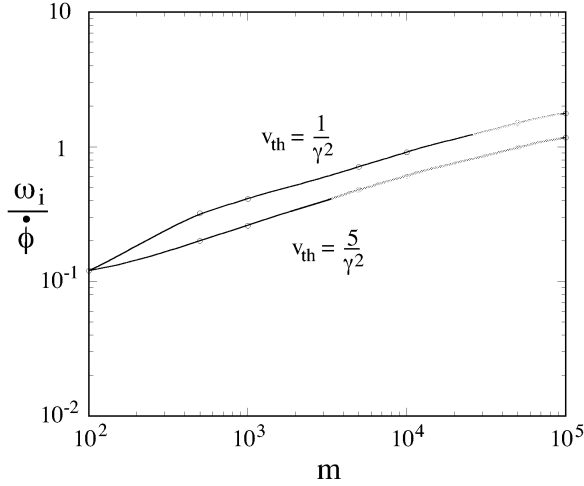}
\caption{Solutions of the dispersion
relation in the presence
of betatron oscillations in
the limit $\bar k_r v_{th} \ll 1$,
$\gamma = 30$, $\zeta = 0.02$.
Points which do not satisfy the inequalities
$m \gg 1$, $m^{2/3} v_{th} < 1$ and
$|\Delta \tilde \omega|^2 < F_0 v_{th}^2$
are plotted in gray.  }
\label{thinbetatron}
\end{figure}

\begin{figure}
\includegraphics[width=3.5in]{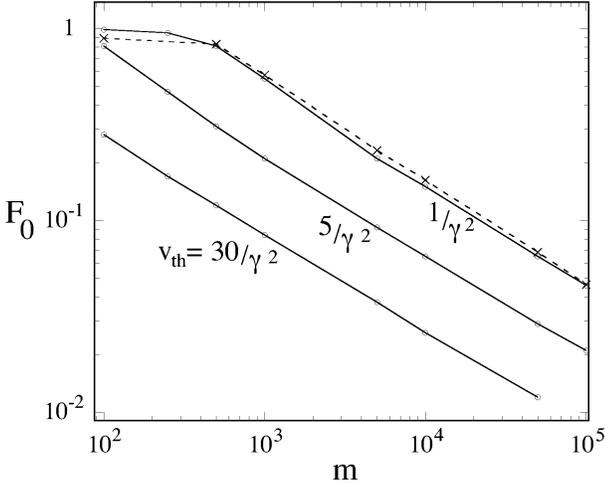}
\caption{$F_0$ as a function of $m$ for $\gamma=30$, $\zeta=0.02$
and various $v_{th}$ and the
squared ratio of the growth rates
from Fig. \ref{growthnobeta}
and Fig. \ref{thinbetatron} (dash-triple-dotted)}
\label{ratioF0}
\end{figure}

In order to determine the lowest
allowed value for $k_r$ and the highest possible growth rate
the full eigenvalue problem has to be solved.
We estimate the result by evaluating Eq. \ref{dpot1}
in the thin approximation again.
Looking at Eq. \ref{dpot1} and replacing the Bessel functions
by their Airy function approximations for the case
$m \gg m_1$ and $m \ll m_2$ we see that the thin approximation
is justified if $\bar k_r v_{th} \ll 1$ and  $m^{2/3} v_{th} \ll 1$.
It starts to fail completely if $m^{2/3} v_{th} \gtrsim 1$, i.e. once
we start integrating over the oscillating and/or the exponentially 
damped/increasing
part of the Airy function, which implies we would like to have
$m^{2/3} | \Delta \tilde \omega | \ll \sqrt{F_0}$ with $|\Delta 
\tilde \omega|^2 \ll F_0 v_{th}^2$
from the previous paragraph. However, for real values of $k_r$ we expect
that the thin approximation will still give us an upper bound of the 
growth rate
because it is easier to maintain coherence if all the radiation is emitted
from the same orbit. With Eq. \ref{drhotthick} we obtain in the limit $\Delta \tilde \omega \ll \gamma^{-2}$
\begin{eqnarray}
1= - \pi~ \zeta~ Z~ \frac{F_0 \gamma^{-2}}
{(\Delta \tilde{\omega} )^2}~,
\label{dispF0}
\end{eqnarray}
The growth rates can be found as before. For $m \gg m_1$ we
obtain
\begin{equation}
\omega_i \simeq {1.083\zeta^{1/2}m^{2/3}\dot\phi\over\gamma} \sqrt{F_0}
\end{equation}
and
\begin{eqnarray}
\nonumber
\omega_i \simeq{\zeta^{1/2}m^{1/2}\dot\phi\over\gamma^{1/2}} \sqrt{F_0}
\end{eqnarray}
for $m \gg m_2$, i.e. there is an additional factor of $\sqrt{F_0}$.
The results for our reference case are plotted in Fig. \ref{thinbetatron} which
were computed numerically. In Fig. \ref{ratioF0} the function $F_0$ is plotted
which we compare with the squared ratio of our new growth rates to the ones
evaluated previously without betatron oscillations.
%%%%%%%%%%%%%%%%%%%%%%%%%%%%%%%%%%%%%%%%%%%%%%
%%%%%%%%%%%%%%%%%%%%%%%%%%%%%%%%%%%%%%%%%%%%%
%
%%%%%%%%%%%%%%%%%%%%%%%%%%%%%%%%%%%%%%%%%%%%%%
%%%%%%%%%%%%%%%%%%%%%%%%%%%%%%%%%%%%%%%%%%%%%

We could also study the effect
of the non-zero thickness alone without
betatron oscillations setting
$F_0=1$ and $F_{n \neq 0}=0$ and solving the
full eigenvalue problem.
Due to the complicated nature of the
dispersion relation we have
not done this yet.
Note that the thin approximation
will suppress certain modes,
e.g. the negative mass instability cannot be
expected to be present with the
   fields having been evaluated at one radius
only, cf. \cite{Briggs1966}.

\section{Spectrum of Coherent Radiation}

\begin{figure}
\includegraphics[width=3.2in]{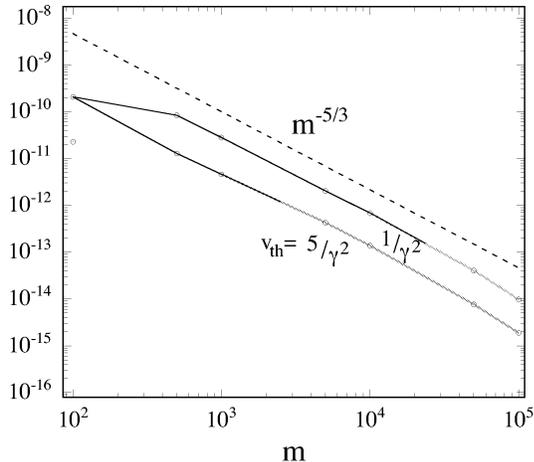}
\caption{Radiated power $(m^{-3} (\omega_i / \dot \phi)^4)$
for $\gamma = 30$ and $\zeta = 0.02$
in arbitrary units. The straight line is proportional
to $m^{-5/3}$ and is shown for comparison.
Points which do not satisfy the inequalities
$m \gg 1$, $m^{2/3} v_{th} < 1$ and
$|\Delta \tilde \omega|^2 < F_0 v_{th}^2$
are plotted in gray. }
\label{radpower}
\end{figure}

Having computed the growth rate
and the saturation amplitude,
the radiated power can now be calculated.
Starting from Eq. (\ref{dIdOmega1}) we now have
\begin{eqnarray}
P_m = {\pi \over 2}  L \omega r_0^4~
\big| \delta J_{\phi 0}\big|^2
\left | \int
\xi d \xi e^{i \bar k_r \xi}
J_m^{\prime}(\omega r_0 \xi) \right |^2~,
\end{eqnarray}
where $\xi \equiv r/r_0$ and the integration is
over the thickness of the layer.
        The Bessel function can be expressed
approximately in term
of an Airy function as done before.
       We take the linear approximation to the
Airy function
as discussed previously, and this gives
\begin{eqnarray}
P_m =  \frac{\pi L \omega}{2} r_0^4 c_2^2
\left ( \frac{2}{m} \right )^{4/3}
\big| \delta J_{\phi 0} \big|^2
\left | \int_{1-v_{th}}^{1+v_{th}}
\xi d \xi e^{i \bar k_r \xi} \right |^2~,
\label{powintegral}
\end{eqnarray}
where $c_2\approx 0.259$.
This is valid for sufficiently big values of
$\gamma$ and low $m$.
The largest values occur for $\bar k_r v_{th} \ll 1$,
where this quantity
is simply $4 v_{th}^2$.
This is enough motivation
for us to work in this limit.
Thus,
\begin{eqnarray}
P_m \le 2\pi L \omega r_0^4 c_2^2 v_{th}^2
     \left | \delta J_{\phi 0}
\right |^2
\left ( \frac{2}{m} \right )^{4/3}~.
\end{eqnarray}
Because we calculated our
growth rates in the thin approximation
for $k \approx k_{\phi}$ it is consistent to use
$\delta \phi = 4 \pi^2 v_{th}
v_{\phi}^{-1} Z r_0 \delta J_{\phi 0}$.
Furthermore, we set $\omega \rightarrow
m \dot \phi$. This is consistent
even for large growth rates since
the exponential growth has stopped.
With our expression for the
saturation amplitude we obtain
\begin{equation}
P_m \le \frac{Lc_2^2v_{\phi}^2 m_e^2}{8 \pi^3 r_0 e^2}
\frac{\gamma^6}{|Z|^2}\left ( \frac{2}{m} \right )^{4/3}
    {1 \over m^3}
\left ( \frac{\omega_i(m)}{\dot \phi} \right )^4~.
\end{equation}
Since the number of particles
$N$ is proportional to $\zeta$ and the growth
rates are proportional
to $\sqrt{\zeta}$ for $m > m_1$
the radiated power scales like $N^2$.
This suggests that the
emitted radiation is coherent.
In Fig. \ref{radpower} we plotted
the radiated power in arbitrary
units having evaluated $F_0$ numerically. For large
$m$ the curve scales as $m^{-5/3}$.
Analytically we obtain with our second approximation for $F_0$
the scaling $m^{-3}  (m^{2/3} / m^{1/4}  )^4=m^{-4/3}$.
With $|Z|^2 \approx 4 c_1^4 \left ( {2}/{m} \right )^{4/3}$
we obtain
\begin{eqnarray} 
P_m \lesssim 3.71 \times 10^{14}
     \gamma^6 m^{-3} \frac{L}{r_0}
\left ( \frac{\omega_i}{\dot \phi} \right )^4
\frac{\rm erg}{\rm s}~.
\end{eqnarray}

%%%%%%%%%%%%%%%%%%%%%%%%%%%%%%%%%%%%%%%%%%%%%%
%%%%%%%%%%%%%%%%%%%%%%%%%%%%%%%%%%%%%%%%%%%%%%%%%%

\subsection{Brightness Temperatures}

          We consider the brightness temperatures
$T_B$ for conditions relevant to the
radio emissions of pulsars.
           Using the Rayleigh-Jeans formula
$B_\nu = 2 k_B T_B (\nu/c)^2$ for
the radiated power per unit area
per sterradian at a frequency $\nu= m \dot{\phi}/2\pi$
gives
\begin{eqnarray}
\nonumber
2k_B T_B ({\nu / c})^2
{\cal A} \Delta \Omega = 2\pi P_m/\dot{\phi}
\\
T_B \lesssim 4.5 \times 10^{21} \frac{\rm K}{\rm m} \cdot
L \gamma^6 m^{-4} \left ( \frac{\omega_i}{\dot \phi} \right )^4
\end{eqnarray}
where $k_B$ is Boltzmann's constant and
${\cal A}=2\pi r_0L$ is the area of the
E-layer.
   The solid angle of the source seen by
a distant observer has been computed in appendix A
and its value is $\Delta \Omega = {4 \pi^2 r_0}/({mL})$.
It is assumed that the angular size of the source is
small such that that radiation from the top and the bottom
emitted at an angle $\theta$ with respect to the normal
is received by the observer at the same position.
For the sample values $\gamma = 1000$,
$\zeta = 0.08$, $v_{th} = 0.04 \gamma^{-2}$,
$L = 100$ km  and  $m=m_1$ our model
predicts a maximum brightness temperature
of $T_B \approx  2 \times 10^{20} {\rm K}$.
According to our results from
previous sections there may be
degeneracy from modes with non-zero
axial wavenumbers $k_z$.
It is reasonable to assume that this will
increase the brightness
temperature by a factor in the order of
$m \tan \psi_{cr}$.
Beaming along the z-axis
may increase the brightness temperature and the observed
frequency even further.

\section{Applications in Accelerator Physics}

The next-generation linear
collider requires a beam
with very short bunches
and low emittance.
     That is,  the beam must
occupy a very small volume in phase space.
     The emittance
of the pre-accelerated
beam is reduced in a damping ring
which is operated with
longer bunches to avoid certain
instabilities.
     The bunch length has to be decreased in a
so-called bunch compressor
before the beam can be injected into
the linear collider.
      A bunch compressor consists of an
accelerating part and
an arc section.
      Since the bunch
lengths of the proposed
linear colliders are in the order of
the wavelength of the synchrotron radiation which is
being radiated in the arc section, instabilities due to
coherent synchrotron have to be taken seriously.
For a design energy of $2$ GeV
and $7 \times 10^{11}$ electrons per 100 $\mu$m our
dimensionless quantities become $\gamma=4000$ and
$\zeta=0.08$ \cite{Raubenheimer}. Our qualitative analysis
of the betatron motion suggests that CSR is suppressed
for a minimum energy spread of
$v_{th} > \zeta^{-1} \gamma^{-2} = 12.5 \gamma^{-2}$.

\section{Discussion and Conclusions}

        This work has studied the
stability of a
collisionless, relativistic,
finite-strength, cylindrical
electron (or positron) layer
by solving the Vlasov and
Maxwell equations.
         This system is of interest to
understanding the high brightness
temperature coherent synchrotron
radio emission of pulsars
and the coherent synchrotron radiation
observed in particle accelerators.
        The considered equilibrium
layers have a finite `temperature'
and therefore a finite radial
thickness.
        The electrons are considered
to move either almost perpendicular
to a uniform external magnetic
field or almost parallel to an
external toroidal magnetic field.
         A short wavelength
instability is found which
causes an exponential growth
an initial perturbation of
the charge and current densities.
         The periodicity of these
enhancements can lead to coherent
emission of synchrotron radiation.
       Neglecting betatron oscillations
we obtain an expression for
the growth rate which is similar
to the one found by Goldreich and Keeley
\cite{GoldreichKeeley1971}
if the thermal energy spread
is sufficiently small.
       The growth rate increases monotonically
approximately as $m^{1/2}$, where $m$ is the
azimuthal mode number which is proportional
to the frequency of the radiation.
       With the radial betatron oscillations included,
the growth rate varies as $m^{1/3}$ over
a significant
range before it begins to decrease.

        We argue that the growth
of the unstable perturbation saturates
when the trapping frequency of
electrons in the wave becomes
comparable to the growth rate.
        Owing to this saturation we can
predict the radiation spectrum
for a given set of parameters.
        For the realistic case including
radial betatron oscillations we
find a radiation spectrum
proportional to $m^{-5/3}$.
       This result
is in rough agreement with observations
of radio pulsars \cite{ManchesterTaylor1977}.
The power is also
proportional to the square of
the number of particles which
indicates that the radiation is
coherent.
    Numerical simulations of electron rings
based on the fully relativistic, electromagnetic
particle-in-cell code OOPIC \cite{schmekel2004b}
recovers  the main scalings
found here.

\acknowledgments

We thank J.T. Rogers, G.H. Hoffstaetter,
and G.S. Bisnovatyi-Kogan for valuable discussions.
This research was partially supported by the
Stewardship Sciences Academic Alliances program
of the National Nuclear Security Administration under
US Department of Energy 
cooperative agreement DE-FC03-02NA00057
and by the National Science
Foundation under contract numbers AST-0307273
and IGPP-1222.

\appendix*
\section{Green's Function}

The Green's function for the potentials give
\begin{eqnarray}
\delta \Phi({\bf r},t)=
\int dt^\prime d^3 r^\prime ~
G({\bf r}-{\bf r^\prime},t-t^\prime) ~
\delta \rho({\bf r}^\prime,t^\prime)~.
\nonumber\\
\delta {\bf A}({\bf r},t)=
\int dt^\prime d^3 r^\prime ~
G({\bf r}-{\bf r^\prime},t-t^\prime) ~
\delta {\bf J}({\bf r}^\prime,t^\prime)~,
\end{eqnarray}
where
$$
\left(\nabla^2 - {\partial^2 \over \partial t^2}\right)
G({\bf r},t)= -4\pi \delta(t)\delta({\bf r})~,
~~ \tilde{G}({\bf k},\omega)=
{4\pi \over  {\bf k}^2-\omega^2  }~,
$$
\begin{eqnarray}
G({\bf r},t)=\frac{4\pi}{(2 \pi)^4} \int_C d\omega \int d^3k~
{\exp(i{\bf k\cdot r}-i\omega t) \over  {\bf k}^2-\omega^2 }~,
\end{eqnarray}
where $\tilde{G}$ is the Fourier transform of the Green's
function. The ``C'' on the integral indicates an
$\omega-$integration parallel to but above the real axis, ${\rm
Im}(\omega)>0$, so as to give the retarded Green's function.

           Because of the assumed dependences
of Eq. (\ref{ansatz}),
we  have for the electric potential,
$$
\delta \Phi_{\omega m k_z}(r) = 2
\int_0^\infty r^\prime dr^\prime
\int_0^\infty \kappa d\kappa \int_0^{2\pi} d\alpha~
~ \delta
\rho_{\omega m k_z}(r^\prime)\big[.. \big]
$$
\begin{equation}
=4\pi \int_0^\infty r^\prime dr^\prime
\int_0^\infty \kappa d\kappa~
{J_m(\kappa r) J_m(\kappa r^\prime)
\over \kappa^2- (\omega^2 - k_z^2) }~
\delta\rho_{\omega m k_z}(r^\prime)~,
\end{equation}
where
$$
\big[.. \big]\equiv{\exp(im\alpha)
J_0\{\kappa [r^2+ (r^\prime)^2-2r
r^\prime
\cos\alpha]^{1/2}\}
\over \kappa^2-(\omega^2 - k_z^2) }~,
$$
where $\kappa^2 \equiv k_x^2+k_y^2$.
Because $\omega$ has a positive imaginary part,
this solution corresponds to the retarded field.
         Also because ${\rm Im}(\omega)>0$, the
$\kappa-$integration can be done by
a contour integration as discussed in \cite{Watson1966}
which gives
\begin{equation}
\delta \Phi_{\omega m k_z}(r)\!=\!
2\pi^2 i \int_0^\infty \!\!\!r^\prime dr^\prime
J_m(kr_<)H_m^{(1)}(kr_>)\delta\rho_{\omega m k_z}(r^\prime),
\label{dpot1}
\end{equation}
where $k\equiv  (\omega^2-k_z^2)^{1/2}$, where $r_<$ ($r_>$)
is the lesser (greater) of $(r,r^\prime)$, and where
$H_m^{(1)}(x)=J_m(x)+iY_m(x)$ is the  Hankel
function of the first kind.
From the Lorentz gauge condition
\begin{equation}
\delta \Psi^{\omega m k_z}(r)= r\delta A_\phi^{\omega m k_z}=
r_0 v_{\phi} \left ( 1 + \Delta \tilde \omega \right )
\delta \Phi^{\omega m k_z}(r)
\label{dpot2}
\end{equation}
Equations (\ref{dpot1}) and (\ref{dpot2}) are useful in subsequent
calculations.

          To determine the total synchrotron radiation
from the E-layer it is sufficient to calculate
$\delta {\bf A}$ at a large distance from the
E-layer.
          We assume that the E-layer has a finite
axial length and exists between $-L/2 \leq z \leq L/2$.
          Thus we evaluate $\delta {\bf A}$ in a spherical
coordinate system ${\bf R}=(R,\theta, \phi)$ at a distance
$R \gg L$.
          The retarded solution is
$$
\delta {\bf A}({\bf R})= {1 \over R}
\int d^3 r^\prime~\delta {\bf J}\bigg({\bf r}^\prime,
t - |{\bf R} -{\bf r}^\prime|\bigg)
=
$$
\begin{equation}
{\exp(i\omega R)\over R}\int\!\!\! d^3 r^\prime~ \delta {\bf
J}(r^\prime)
\exp \bigg[im \phi^\prime +ik_z z^\prime
-i\omega (t +\hat{\bf R}\cdot {\bf r}^\prime)\bigg],
\end{equation}
(see, e.g. ch. 9 of \cite{LandauLifshitz1962}).
          The source point is at
$(x^\prime=r^\prime\cos\phi^\prime,~ y^\prime=r^\prime \sin
\phi^\prime,~ z^\prime)$.
          The observation point is taken to be at
$(x=0,~y=R\sin \theta,~ z=R\cos\theta)$.
          Consequently, $\hat{\bf R}\cdot {\bf r}^\prime=
r^\prime \sin\theta \sin\phi^\prime +z^\prime \cos\theta$.
         The phase factor $\exp(i\omega R)$ does
not affect the radiated power and is henceforth dropped.

           For the cases where $\delta J_\phi$ is the dominant
component of the current-density perturbation we have
$$
\bigg[\delta A_x^\omega,~\delta A_y^\omega\bigg]=
{S(\theta)\over R}\int r^\prime d r^\prime d\phi^\prime~
\bigg[-\sin\phi^\prime, \cos \phi^\prime\bigg] \times
$$
\begin{equation}
\delta J_\phi(r^\prime)~
\exp(im\phi^\prime-i\omega r^\prime \sin \theta
\sin\phi^\prime)~,
\label{dAomega1}
\end{equation}
where
\begin{eqnarray}
S(\theta) \equiv L~ {\sin[(k_z-\omega \cos \theta)L/2]
\over (k_z-\omega \cos \theta)L/2}
\end{eqnarray}
is a structure function accounting for the finite
axial length of the E-layer, and $\omega$ superscript
indicates $\omega=m\dot{\phi}$.
          Carrying out the $\phi^\prime$ integration in
equation (\ref{dAomega1}) gives
\begin{equation}
\bigg[\delta A_x^\omega,~\delta A_y^\omega \bigg]
={S(\theta)\over R}\int r^\prime d r^\prime~
\delta J_\phi(r^\prime)\big[..\big]
\label{dAomega}
\end{equation}
where
$$
\big[..\big]\equiv\left[ i J_m^\prime
(\omega r^\prime \sin \theta),~
{m \over \omega r^\prime \sin \theta}
J_m(\omega r^\prime \sin \theta)\right]
$$
and where the prime on the Bessel function
indicates its derivative with respect to its
argument.
           The radiated power per unit solid angle is
\begin{eqnarray}
{d P_\omega \over d \Omega}
={R^2 \over 8\pi}|\delta {\bf B}^\omega|^2
= {R^2 \over 8\pi}|{\bf k} \times \delta {\bf A}^\omega |^2 =
        \nonumber\\
{R^2 \omega^2 \over 8\pi} \left( |\delta A_x^\omega|^2  +
\cos^2 \theta |\delta A_y^\omega|^2\right)~,
\label{dIdOmega1}
\end{eqnarray}
where ${\bf k} \equiv \omega \hat{\bf R}$ is the
far field  wavevector.

          For a radially thin E-layer, $(\Delta r/r_0)^2 \ll 1$,
equations (\ref{dAomega}) and (\ref{dIdOmega1}) give
\begin{eqnarray}
\nonumber
{d P_\omega \over d \Omega}
={S^2(\theta) \over 8\pi}\left|
\int r^\prime dr^\prime~\delta J_\phi(r^\prime)
\omega J_m^\prime(\omega r_0\sin\theta) \right |^2 +
\\
{S^2(\theta) \over 8\pi} \left | \int r^\prime dr^\prime~\delta
J_\phi(r^\prime) {m~{\rm cot}~\theta \over r_0} J_m(\omega
r_0\sin\theta)\right |^2~.
\label{dIdOmega}
\end{eqnarray}
The factor within the curly brackets is the same as
that for the radiation pattern of a single charged particle
(see ch. 9 of \cite{LandauLifshitz1962}).

          The factor $S^2(\theta)$
in Eq. (\ref{dIdOmega})  tightly
constrains
the radiation to be in the direction
$\theta_* = \cos^{-1}(k_z/\omega)$ if the  angular width of
$S^2(\theta)$, the half-power half-width
$\Delta \theta_{1/2} \approx \pi /(\omega L)$, is
small compared with the angular spread of the single
particle synchrotron radiation, $1/\gamma$,
which is the angular width due to the Bessel
function terms in Eq. (\ref{dIdOmega}).
         This corresponds to E-layers with $L \gg \pi \gamma/\omega
=\pi r_0 \gamma/m$.  For $L \sim r_0$, we
need $m  \gg \pi \gamma$, which is satisfied by
the spectra discussed later in \S 8.
In this case, Eq. (\ref{dIdOmega})
can be integrated over
the solid angle to give
$$
P_\omega = {\pi  L\sin\theta_* \over 2 \omega} \left \{ \left|
\int r^\prime dr^\prime~\delta J_\phi(r^\prime)
\omega J_m^\prime( \omega r_0\sin\theta_*) \right |^2 + \right .
$$
\begin{equation}
\left . \left| \int r^\prime dr^\prime~\delta J_\phi(r^\prime)
{m~{\rm cot}~\theta_* \over r_0} J_m(\omega r_0
\sin\theta_*) \right |^2 \right \}~.
\label{Iomega}
\end{equation}

One limit of  interest of Eq. \ref{Iomega} is that where
$k_z=0$ so that $\theta_*=\pi/2$ and
\begin{eqnarray}
P_m
={\pi m v_\phi L  \over 2r_0 }\left|
\int r^\prime dr^\prime~\delta J_\phi(r^\prime)
J_m^\prime(\omega r_0) \right|^2~,
\label{Imeq}
\end{eqnarray}
where we have set $\omega \rightarrow m \dot{\phi}$.
         The total radiated power is $P = \sum_m P_m$.


\begin{thebibliography}{9}

\bibitem[Gold (1968)]{Gold1968} Gold, T. 1968, Nature, 218, 731

\bibitem[Gold (1969)]{Gold1969} Gold, T. 1969, Nature, 221, 25

\bibitem[Goldreich \& Keeley (1971)]{GoldreichKeeley1971}
Goldreich, P., \& Keeley, D.A. 1971, ApJ, 170, 463

\bibitem[Manchester \& Taylor (1977)]{ManchesterTaylor1977}
Manchester, R.N., \& Taylor, J.H. 1977, {\it Pulsars},
(Freeman \& Co.: San Francisco)

\bibitem[Melrose (1991)]{Melrose1991} Melrose, D. B. 1991,
Ann. Rev. Astron. \& Astrophys., 29, 31

\bibitem[Bisnovatyi-Kogan \& Lovelace (1995) ]{GBKRL95}
Bisnovatyi-Kogan, G.S., \& Lovelace, R.V.E. 1995, A\&A, 296,
L17

\bibitem[Byrd (2002)]{Byrd2002} J.M. Byrd,
Phys. Rev. Lett. {\bf 89}, 224801, (2002)

\bibitem[Kuske (2003)]{Kuske2003} M. Abo-Bakr et al.,
Phys. Rev. Lett. {\bf 90}, 094801 (2003)

\bibitem[Loos (2002)]{Loos2002}
Loos, H. et al., Proceedings of the EPAC 2002, Paris, France

\bibitem[Byrd (2003)]{Byrd2003} F. Sannibale et al., Proceedings
of the 2003 Particle Accelerator Conference, May 2003, Portland
Oregon (edited by J.~Chew, P.~Lucas and S.~Webber, IEEE, Piscataway, New Jersey, 2003)

\bibitem[Heifets \& Stupakov (2001)]{HeifetsStupakov2001}
Heifets, S. \& Stupakov, G. 2001, SLAC Technical Report No.
SLAC-PUB-8761

\bibitem[Stupakov \& Heifets (2002)]{Stupakov2002}
Stupakov, G. \& Heifets, S. 2002, Phys. Rev. AB, 5, 54402

\bibitem[Heifets (2001)]{Heifets2001} Heifets, S. 2001,
SLAC Technical Report No. SLAC-PUB-9054

\bibitem[Uhm et al. (1985)]{Uhm1985}
Uhm, H.S., Davidson, R.C., \& Petillo, J.J.
1985, Phys. Fluids, 28, 2537

\bibitem[Venturini \& Warnock (2002)]{Venturini2003}
Venturini, M. \& Warnock, R. 2002, Phys. Rev. Lett., 89, 224802

\bibitem[Christofilos (1958)]{Christofilos1958}
Christofilos, N. 1958, in {\it Proc. Second U.N.
International Conference on the Peaceful Uses
of Atomic Energy}, Geneva, Vol. 32, p. 279

\bibitem[Goldreich \& Julian (1969)]{Julian1969}
Goldreich, P., \& Julian, W.H. 1969, ApJ, 157, 869

\bibitem[Arons (2004)]{Arons2004}
Arons, J., Advances in Space Research 33 (2004), 466-474

\bibitem[Davidson (1974)]{Davidson1974}  Davidson, R.C. 1974,
{\it Theory of Nonneutral Plasmas} (W.A. Benjamin: New York)

\bibitem[Landau (1946)]{Landau1946}  Landau, L.D. 1946,
J. Phys. U.S.S.R., {\bf 10}, 25

\bibitem[Landau \& Lifshitz (1962)]{LandauLifshitz1962}
Landau, L.D., \& Lifshitz, E.M. 1962,
{\it The Classical Theory of Fields} (Pergamon Press: London)

\bibitem[Kolomenskii \& Lebedev (1959)]{Kolomenskii1959}
Kolomenskii, A.A., \& Lebedev, A.N. 1959,
in {\it Proc. of International Conference on High
Energy Accelerators and Instrumentation} (Geneva: CERN), p. 115

\bibitem[Lawson (1988)]{Lawson1988}
Lawson, J.D. 1988, {\it The Physics of Charged Particle
Beams}, (Clarendon Press: Oxford)

\bibitem[Nielson et al. (1959)]{Nielson1959}
Nielson, C.E., Sessler, A.M., \& Symon, K.R. 1959,
in {\it Proc. of International Conference on High
Energy Accelerators and Instrumentation} (Geneva: CERN),
p. 239

\bibitem[Montgomery \& Tidman (1964)]{Montgomery1964}
Montgomery, D.C., \& Tidman, D.A. 1964, {Plasma
Kinetic Theory}, (McGraw-Hill: New York)

\bibitem{Briggs1966} R.~J.~Briggs and V.~K.~Neil,
Plasma Physics, Vol. 9, pp. 209-227 (1967)

\bibitem{Raubenheimer}
Tor Raubenheimer, SLAC NLC Note 2

\bibitem{chao} A.W.~Chao, R.D.~Ruth, Particle Accelerators, 1985, Vol. 16, pp.
201-216, Gordon and Breach

\bibitem{schmekel} B.~S.~Schmekel, G.~H.~Hoffstaetter and J.~T.~Rogers,
%``Investigation Of The Flat-Beam Model Of The Beam-Beam Interaction,''
Phys.\ Rev.\ ST Accel.\ Beams {\bf 6}, 104403 (2003)

\bibitem{schmekel2004b} Schmekel, B.S., to be submitted

\bibitem[Abramowitz \& Stegun (1965)]{Abramowitz1965}
Abramowitz, M., \& Stegun, I.A. 1965,
{\it Handbook of Mathematical Functions},
(Dover: New York), p. 367

\bibitem[Teukolsky et al. (1989)]{Teukolsky1989}
Press, W.H., Flannery, B.P., Teukolsky, S.A.,
\& Vetterling, W.T. 1989, {\it Numerical
Recipes,} (Cambridge University Press: Cambridge)

\bibitem[Watson (1966)]{Watson1966}
Watson, G.N. 1966, {\it A Treatise on the Theory
of Bessel Functions}, pp. 428-429, (Cambridge University Press:
Cambridge)

\bibitem[Botten et al. (1983)]{Botten1983}
Botten, L.C., Craig, M.S. and McPhedran, R.C.,
Computer Physics Communication {\bf 29}, 245-259 (1983)

\end{thebibliography}
\end{document}